\pdfoutput=1
\documentclass[12pt, oneside]{article}   	
\usepackage[svgnames]{xcolor}
\usepackage[colorlinks]{hyperref}

\usepackage[total={6in, 9in},top=1in,left=1in]{geometry}
\usepackage{graphicx}				
\usepackage{caption} 
\usepackage{amssymb}
\usepackage{hyperref}
\usepackage{bm}
\usepackage{amsmath}
\usepackage{bbm}
\usepackage{amsthm}	
\usepackage{natbib}
\usepackage{tikz}
\usepackage{capt-of}
\usepackage{extarrows}
\usepackage{enumerate}
\usepackage{natbib}
\usepackage{booktabs}
\usepackage{multirow}

\newtheorem{prop}{Proposition}
\newtheorem{ass}{Assumption}

\newtheorem{lemma}{Lemma}

\newtheorem{example}{Example}

\newcommand{\abs}[1]{\left\lvert #1 \right\rvert}

\newcommand{\under}[1]{\underline{#1\mkern-2mu}\mkern2mu }

\hypersetup{
	citecolor=DarkBlue,
	bookmarksnumbered=true,
	urlcolor=Indigo,linkcolor=blue
}

\setcitestyle{authoryear,open={(},close={)}}

\usepackage{setspace}
\newcommand\Xz{\tilde{X}}
\newcommand\xz{ \tilde{x}}
\newcommand\p{p_W}

\title{Partial Identification of Binary Choice Models with Misreported Outcomes\footnote{We are grateful to Jason Blevins, Robert de Jong, Xavier D'Haultfoeuille, Wayne Yuan Gao, Marc Henry, Keisuke Hirano,  Sung Jae Jun,  D\'esir\'e K\'edagni, Joris Pinkse,  and Takuya Ura for their valuable comments and suggestions.}}
\author{ 
Orville Mondal\thanks{Economist, Bates White Economic Consulting. Email: orville.dm@gmail.com}  \ and \ Rui Wang\thanks{Department of Economics, The Ohio State University. Email: wang.16498@osu.edu }   
}

\date{\today}							

\begin{document}

\sloppy

\global\long\def\a{\alpha}%
 
\global\long\def\b{\beta}%
 
\global\long\def\g{\gamma}%
 
\global\long\def\d{\delta}%
 
\global\long\def\e{\epsilon}%
 
\global\long\def\l{\lambda}%
 
\global\long\def\t{\theta}%
 
\global\long\def\o{\omega}%
 
\global\long\def\s{\sigma}%

\global\long\def\G{\Gamma}%
 
\global\long\def\D{\Delta}%
 
\global\long\def\L{\Lambda}%
 
\global\long\def\T{\Theta}%
 
\global\long\def\O{\Omega}%
 
\global\long\def\R{\mathbb{R}}%
 
\global\long\def\N{\mathbb{N}}%
 
\global\long\def\Q{\mathbb{Q}}%
 
\global\long\def\I{\mathbb{I}}%
 
\global\long\def\P{\mathbb{P}}%
 
\global\long\def\E{\mathbb{E}}%
\global\long\def\B{\mathbb{\mathbb{B}}}%
\global\long\def\S{\mathbb{\mathbb{S}}}%
\global\long\def\V{\mathbb{\mathbb{V}}\text{ar}}%
 
\global\long\def\GG{\mathbb{G}}%
\global\long\def\TT{\mathbb{T}}%

\global\long\def\X{{\bf X}}%
\global\long\def\cX{\mathscr{X}}%
 
\global\long\def\cY{\mathscr{Y}}%
 
\global\long\def\cA{\mathscr{A}}%
 
\global\long\def\cB{\mathscr{B}}%
\global\long\def\cF{\mathscr{F}}%
 
\global\long\def\cM{\mathscr{M}}%
\global\long\def\cN{\mathcal{N}}%
\global\long\def\cG{\mathcal{G}}%
\global\long\def\cC{\mathcal{C}}%
\global\long\def\sp{\,}%

\global\long\def\es{\emptyset}%
 
\global\long\def\mc#1{\mathscr{#1}}%
 
\global\long\def\ind{\mathbf{\mathbbm1}}%
\global\long\def\indep{\perp}%

\global\long\def\any{\forall}%
 
\global\long\def\ex{\exists}%
 
 
\global\long\def\cd{\cdot}%
 
\global\long\def\Dif{\nabla}%
 
\global\long\def\imp{\Rightarrow}%
 
\global\long\def\iff{\Leftrightarrow}%

\global\long\def\up{\uparrow}%
 
\global\long\def\down{\downarrow}%
 
\global\long\def\arrow{\rightarrow}%
 
\global\long\def\rlarrow{\leftrightarrow}%
 
\global\long\def\lrarrow{\leftrightarrow}%

\global\long\def\abs#1{\left|#1\right|}%
 
\global\long\def\norm#1{\left\Vert #1\right\Vert }%
 
\global\long\def\rest#1{\left.#1\right|}%

\global\long\def\bracket#1#2{\left\langle #1\middle\vert#2\right\rangle }%
 
\global\long\def\sandvich#1#2#3{\left\langle #1\middle\vert#2\middle\vert#3\right\rangle }%
 
\global\long\def\turd#1{\frac{#1}{3}}%
 
\global\long\def\ellipsis{\textellipsis}%
 
\global\long\def\sand#1{\left\lceil #1\right\vert }%
 
\global\long\def\wich#1{\left\vert #1\right\rfloor }%
 
\global\long\def\sandwich#1#2#3{\left\lceil #1\middle\vert#2\middle\vert#3\right\rfloor }%

\global\long\def\abs#1{\left|#1\right|}%
 
\global\long\def\norm#1{\left\Vert #1\right\Vert }%
 
\global\long\def\rest#1{\left.#1\right|}%
 
\global\long\def\inprod#1{\left\langle #1\right\rangle }%
 
\global\long\def\ol#1{\overline{#1}}%
 
\global\long\def\ul#1{\underline{#1}}%
 
\global\long\def\td#1{\tilde{#1}}%
\global\long\def\bs#1{\boldsymbol{#1}}%

\global\long\def\upto{\nearrow}%
 
\global\long\def\downto{\searrow}%
 
\global\long\def\pto{\overset{p}{\longrightarrow}}%
 
\global\long\def\dto{\overset{d}{\longrightarrow}}%
 
\global\long\def\asto{\overset{a.s.}{\longrightarrow}}%

\setlength{\abovedisplayskip}{6pt} \setlength{\belowdisplayskip}{6pt}

\maketitle
\begin{abstract}


This paper provides partial identification of various binary choice models with misreported dependent variables. We propose two distinct approaches by exploiting different instrumental variables respectively. In the first approach, the instrument is assumed to only affect the true dependent variable but not misreporting probabilities. The second approach uses an instrument that influences misreporting probabilities monotonically while having no effect on the true dependent variable. Moreover, we derive identification results under additional restrictions on misreporting, including bounded/monotone misreporting probabilities. We use simulations to demonstrate the robust performance of our approaches, and apply the method to study educational attainment.




\end{abstract}

\newpage

\section{Introduction}

This paper provides partial identification of various binary choice models when the dependent variable is potentially misreported. Binary choice models have been widely used in empirical applications such as analyzing participation in social programs, employment status, and educational attainment. However, many applications rely on survey data such as the Survey of Income and Program Participation (SIPP) and the Current Population Survey (CPS), where the binary outcome variable may be misreported or misclassified in survey data due to interviewer or respondent errors. The problem of misreporting is well documented and several studies show that the misreporting probabilities can be significant. For example, \cite*{meyer2020} show that the
probability of misreporting participation in a food stamp program can range from 23\% in the SIPP to 50\% in the CPS.


Numerous studies have examined the bias introduced by misreporting across various  econometric models \citep*{aigner1973, bollinger1997, kane1999, davern2009, nguimkeu2019}. Regarding a binary choice model, \cite{meyer2017} show that misreporting in the binary dependent variable can lead to significant biases in parametric estimators.  While misreporting might be pervasive in some widely used datasets, these datasets may remain valuable sources of information, often with no appropriate substitute. It is therefore vital to investigate what can still be learned from the contaminated data.

Tackling misreporting issues can be challenging. Firstly, misreporting in a binary variable involves non-classical measurement errors, as the measurement error is always negatively correlated with the true outcome. Moreover, misreporting stem from unobserved incentives among respondents; for example, people who benefit from a food stamp program may conceal their participation out of a sense of shame. As such, misreporting probabilities can depend on observed characteristics in an unknown way.

This paper introduces two different approaches to identify various binary choice models with a potentially misreported dependent variable, including parametric, semiparametric, and panel binary choice models. With potential misreporting in the dependent variable, conventional approaches for binary choice models do not apply, as the true dependent variable is not observed, and the conditional expectation of the true dependent variable is not identified. Our identification strategy derives bounds for the conditional expectation of the true outcome given covariates, by exploiting variation in different instruments. Given the bounds, we derive partial identification for binary choice models by characterizing conditional moment inequalities.


In our first approach, the instrument is assumed to only affect the true outcome while not influencing misreporting probabilities. In the example of program participation, such as job training program, this instrument can be randomly assigned eligibility for the program. This instrument affects the true participation of the program, but it is unlikely to affect misreporting, given its random nature. 
The second approach uses an instrument that only affects misreporting probabilities monotonically, but does not influence the true outcome. Examples of such a variable could include interview-relevant variables such as an interviewer's evaluation of respondents' accuracy or interview styles in survey data, including in-person, phone, or email interviews. Individuals are more likely to provide truthful responses during in-person interviews than during email interviews. 

We derive partial identification results using each instrument. The strategy involves deriving bounds on misreporting probabilities through the variation in each instrument, thereby establishing bounds on the conditional expectation of the true outcome. Furthermore, we explore the identifying power of each instrument with additional restrictions on misreporting, including one-sided misreporting, bounded misreporting probabilities, and monotone misreporting probabilities. These restrictions can potentially provide more informative bounds for misreporting probabilities and thus tighten the bounds for the conditional expectation of the true outcome.



Our approach accommodates various binary choice models and allows for flexible misreporting processes.  The identification strategy is applicable to binary choice models under full independence and distributional assumptions, or median independence restrictions, as well as panel models with conditional homogeneity conditions. Additionally, we allow for heterogeneous misreporting probabilities, which can depend on observed characteristics arbitrarily. This flexibility has practical value; for example, \cite{bollinger1997, bollinger2001} demonstrate that misreporting probabilities are correlated with participants' characteristics such as demographic characteristics and family income. Furthermore, we do not assume a parametric model for the misreporting process (such as a linear index model) and permit arbitrary dependence between the true outcome and the misreporting process. 

We characterize partial identification for various binary choice models using conditional moment inequalities. Through simulations, we evaluate the finite sample performance of our method, taking the semiparametric model as an illustrative example.   For comparison, we also implement the maximum likelihood estimation method studied in \cite*{hausman1998}, which assumes constant misreporting probabilities and distributional assumptions. The results demonstrate the robustness of our approaches with respect to heterogeneous misreporting probabilities and parametric assumptions. As an empirical illustration, we apply our method to study a binary choice model of educational attainment using the data from National Longitudinal Surveys in 1976. 


In the extension, we examine the joint identifying power of the two instrumental variables together. The two instruments jointly provide a new channel for identification, introducing additional restrictions on misreporting probabilities. This result leads to more informative bounds on the conditional expectation of the true outcome compared to intersecting the bounds obtained by using each instrument separately.

\subsection{Related Literature}

This paper directly contributes to the line of literature on binary choice models with misreported dependent variables. See \cite*{chen2011nonlinear} for a comprehensive survey of nonlinear models with measurement errors, including binary choice models. The literature on binary choice models dates back to \cite{chamberlain1980},  \cite{manski1985}, and \cite{manski1987}. 
When the dependent variable is misreported,  \cite*{hausman1998} proposes a modified maximum likelihood estimator for binary choice models to correct potential misreporting, and \cite{abrevaya1999semiparametric} explores a more general linear index model. These studies assume homogeneous misreporting probabilities, where misreporting rates are constant regardless of values of covariates. \cite{bollinger1997} allows misreporting rates to depend on covariates in a Probit model, imposing parametric assumptions for both the true binary choice model and misreporting processes. 
\cite{lewbel2000} studies semiparametric identification of binary choice models by using a continuous instrumental variable that only affects the true outcome but is independent of misreporting rates. 
In a more recent paper, \cite{meyer2017} proposes different parametric estimators, relying on a parametric model for misreporting processes or the availability of validation data. Our paper allows for heterogeneous misreporting probabilities, which can depend on covariates in an arbitrary way.
Furthermore, we explore the identifying power of discrete instruments (along with additional restrictions on misreporting) and the results are valid even if the instrument only take two values.

Our work also relates to a large body of literature on various models with misreported regressors.  \cite{mahajan2003misclassified} investigates misclassified regressors in binary choice models. Several studies explore regression models with misreported regressors under homogeneous misreporting probabilities, including \cite{aigner1973}, \cite{bollinger1996}, \cite{frazis2003}.   \cite{mahajan2006}, \cite{lewbel2007}, \cite{ditraglia2019} allow misreporting probabilities to be covariates dependent and achieve identification by using a binary instrument for the true regressor.  \cite*{chen2005measurement}, \cite{chen2006identification}, and \cite*{chen2008nonparametric} provide nonparametric identification without instrumental variables. They achieve identification by exploiting auxiliary data, two samples, and higher-order moments, respectively. 
Additionally, \cite{hu2008} and \cite{molinari2008} explore misclassification in a general discrete regressor, and \cite{hu2008instrumental} and \cite*{hu2022identification} extend the study to a continuous regressor with nonclassical measurement errors. These papers investigate different frameworks with misreported regressors, and thus their assumptions and identification approaches/results differ from those in our paper. In a more closely related study, \cite*{nguimkeu2019} uses two instruments jointly for identification--one for the true regressor and the other for misreporting. They obtain point identification under one-sided misreporting and parametric structures on misreporting, while our paper allows for two-sided misreporting and nonparametric structures on misreporting processes. 

The above literature focuses on homogeneous effects of the regressor given covariates in a regression setting. Numerous papers study heterogeneous treatment effects with a misreported treatment
 such as \cite{kreider2007}, \cite*{kreider2012}, \cite*{battistin2014}, \cite*{calvi2017}, and \cite{ura2018}.  Their approaches either exploit a repeated measurement, auxiliary administrative data to restrict misreporting errors, or an instrumental variable related to the true treatment. This literature also studies different framework and requires different assumptions to identify heterogeneous treatment effects from our paper.\footnote{Our approach can be potentially applied to study heterogeneous treatment effects, while it still requires substantial work to explore how to combine our approach with additional assumptions on the instrument for identification. } Additionally, these works exploit the instrument for treatment, while we also explore the identifying power of the instrument for misreporting. 
 More differently, \cite{horowitz1995identification}, \cite{kreider2008inferring}, and \cite{kreider2011identification} study identification with corrupted data under a mixing framework on data errors.


The rest of this paper is organized as follows. Section \ref{sec:model} presents various binary choice models and identification approaches using two different instrument along with additional assumptions on misreporting. Section \ref{sec:cmi} characterizes conditional moment inequalities for model parameters, and Section \ref{sec:sim} examines the finite sample performance via simulations. Section \ref{sec:emp} studies the application of educational attainment. Section \ref{sec:exte} explores an extension. We conclude with Section \ref{sec:conc}.

\section{Model and Identification} \label{sec:model}

The analysis studies the identification of various binary choice models with potential misreporting (or misclassification) in the binary dependent variable.
Let $Y_i^{*}\in\{0 ,1\}$ denote the true binary dependent variable, and $Y_i\in \{0, 1\}$ denote the observed  variable which may be subject to misreporting.  Let $X_i\in \mathcal{X}$ denotes a vector of observed covariate, which is relevant to the true variable $Y_i^{*}$ and can also affect misreporting probabilities. 

When there is potential misreporting in the variable $Y_i^{*}$, the standard identification results do not apply, as the true variable $Y_i^{*}$ is not observed and the conditional probability $p^{*}(x):=\Pr(Y_i^{*}=1 \mid x)$ is not identified. 
Our identification strategy is to establish bounds $[L(x), U(x)]$ for the true conditional choice probability using observed variables $(X_i, Y_i)$ and exploiting the availability of various instruments. The bounds $[L(x), U(x)]$ will depend only on observed variables, so they are identified from data. Based on the bounds, we can characterize partial identification for several binary choice models.
   
Before introducing the identification approach, we first present various binary choice models to illustrate how the bounds on $p^{*}(x)$ can be exploited to derive partial identification results in various models.
 
\begin{example}[Parametric Binary Choice Model] \label{example:par}
Consider the following model for the true dependent variable $Y_i^{*}$:
\[
Y_i^{*} = \ind\{\e_i \leq X_i'\b_0\},
\]
where $\e_i$ is independent of $X_i$ and follow a known distribution: $\e_i \mid X \sim F_{\e}(\cdot)$.\footnote{Common examples for parametric binary choice models include Probit and Logit models.} Under this structure, the true conditional choice probability is given as
\[
p^{*}(x)= F_{\e}(x'\b_0).
\]

We are interested in identifying and estimating the parameter $\b_0$. 
Given the derived bounds $p^{*}(x)\in [L(x), U(x)]$ in Section \ref{sec:inde}, the identifying condition for the true parameter $\b_0$ is characterized by the following conditional moment inequality: for any $x$, 
\[
L(x) \leq F_{\e}(x'\b_0) \leq U(x).
\]

\end{example}

\begin{example}[Semiparametric Binary Choice Model] \label{example:semi}
The true dependent variable $Y_i^{*}$ is given as:
\[
Y_i^{*} = \ind\{\e_i \leq X_i'\b_0\},
\]
where $\e_i$ satisfies the mean independence assumption $E[\e_i \mid X_i]=0$, while the distribution of $\e_i$ is left unknown. When there is no misreporting in the variable $Y_i^{*}$, \cite{manski1985} derives the following identifying condition for $\b_0$:
\[ \begin{aligned}
 x'\b_0 \geq 0 &\Longrightarrow p^{*}(x) - 0.5 \geq 0, \\
 x'\b_0 \leq 0 &\Longrightarrow p^{*}(x) - 0.5 \leq 0. \\
\end{aligned}
\]

However, the probability $p^{*}(x)$ is no longer identified when the true outcome $Y_i^{*}$ is not observed. Our approach dervies bounds for the conditional choice probability $p^{*}(x)\in [L(x), U(x)]$, yielding the following identifying condition:
\[
\begin{aligned} 
x'\b_0 \geq 0\Longrightarrow p^{*}(x)-0.5\geq 0 \Longrightarrow U(x)-0.5\geq 0, \\
x'\b_0 \leq 0 \Longrightarrow p^{*}(x)-0.5\leq 0 \Longrightarrow L(x)-0.5\leq 0.
\end{aligned}
\]

\end{example}

\begin{example}[Panel Binary Choice Model] \label{example:panel}

Consider the following binary choice model for $Y_{it}^{*}$ with panel data structure: 
\[
Y_{it}^{*} = \ind\{\e_{it}+\a_i \leq X_{it}'\b_0\},
\]
where $\e_{it}$ satisfies the conditional stationarity (homogeneity) assumption in \cite{manski1987}: $\e_{it}\mid \a_i, X_{is}, X_{it} \sim \e_{is}\mid \a_i, X_{is}, X_{it}$. Similar to Example \ref{example:semi}, given the bounds on $p^{*}_t(x):=\Pr(Y_{it}^{*}=1 \mid X_{ist}=x)\in [L_t(x), U_t(x)]$ where $X_{ist}:=(X_{is}, X_{it})$, it has the following implication: 
\[ \begin{aligned}
(x_t-x_s)'\b_0 \geq 0 &\Longrightarrow p^{*}_t(x)  \geq p^{*}_s(x) \Longrightarrow U_t(x)\geq L_s(x), \\
 (x_t-x_s)'\b_0 \leq 0 &\Longrightarrow p^{*}_t(x) \leq p^{*}_s(x) \Longrightarrow L_t(x)\leq U_s(x). \\
\end{aligned}
\]

\end{example}

~

Although our paper mainly focuses on binary choice models, the proposed approach can be potentially applied to estimate treatment effects with misreported treatment.

\begin{example}[Local Average Treatment Effects]

Suppose that we are interested in estimating the causal effects of the true treatment $T^{*}\in \{0, 1\}$ on the outcome $Y$. The true treatment $T^{*}$ can be endogenous, so a binary instrument $Z\in\{0, 1\}$ is used to address endogeneity. Under the assumptions on instrument $Z$ introduced in \cite{imbens1994identification}, the local average treatment effect (LATE) is given as 
\[
LATE =\frac{E[Y\mid Z=1]-E[Y\mid Z=0]} {E[T^{*} \mid Z=1]-E[T^{*}\mid Z=0]}.
\] 

We consider that the true treatment $T^{*}$ is not observed, but instead we only observe a reported treatment $T\in\{0, 1\}$ which could be subject to misreporting. Our approach can bound the true conditional probability $p^{*}(z):= E[T^{*}\mid Z=z] \in [L(z), U(z)]$, and thus can also bound LATE.\footnote{The heterogeneous treatment effects framework introduces additional assumptions on instrument $Z$ with $(T^*, Y)$ to address endogeneity. Our approach only imposes assumptions between instruments $(Z, W)$ and variables $(T^{*}, T)$ (excluding $Y$) to address misreporting issues. Therefore, under this framework, we need to combine all these assumptions jointly to identify treatment effects with misreported treatment. }

\end{example}

\subsection{Identification}\label{sec:inde}

We now present our identification approaches to establish bounds on the true conditional probability $p^{*}(x)=\Pr(Y_i^{*}=1\mid x)$. Let $p(x)=\Pr(Y_i=1\mid x)$ denote the reported probability of $Y_i=1$ given $X_i=x$, which is identified from the data. The reported probability $p(x)$ depends on two components: the true probability and misreporting probabilities. Therefore, it is essential to distinguish between these two components to identify the true probability from the reported probability. 
%

We introduce two different approaches to identify the true probability $p^{*}(x)$ by exploiting different exclusion restrictions. In the first approach, we use a discrete instrument $Z_i\in\mathcal{Z}:=\{z_1, z_2,..., z_k\}$ that only affects the true probability $p^{*}(x)$ but does not affect misreporting probabilities. The other approach uses a discrete instrument $W_i\in \mathcal{W}:=\{w_1, w_2,..., w_l\}$ that only affects misreporting probabilities but not the true probability $p^{*}(x)$. We first study the identifying power of each individual instrument and then discuss how the two instruments can jointly identify the true probability $p^{*}(x)$ in the extension. 


For simplicity of notation, we suppress subscript $i$ for random variables in the following analysis. The following graph describes the relationship among all variables.


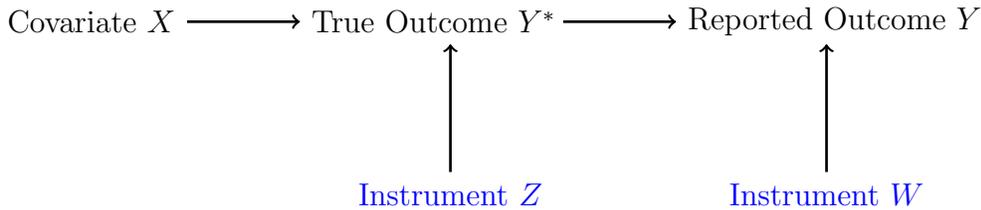
\begin{figure}[h!]
\caption{Relationship among Variables}
 \label{figure}
\begin{tikzpicture}[line width=1pt] 
 \centering
\usetikzlibrary{decorations.pathreplacing}
\draw[->] (1,0)--(2.5,0);
\node[left] at (1,0) {Covariate $X$};
\node[right] at (2.5,0) { True Outcome $Y^{*}$};
\draw[->] (6,0)--(7.5,0);
\node[right] at (7.5,0) {Reported Outcome $Y$};
\draw[<-] (4.5,-0.3)--(4.5,-2);
\node[below] at (4.5,-2) {{\color{blue} Instrument $Z$}};
\draw[<-] (9.5,-0.3)--(9.5,-2);
\node[below] at (9.5,-2) {{\color{blue}Instrument $W$}};
\end{tikzpicture}
\end{figure}


Figure \ref{figure} summarizes our model and main identification strategies. The objective is to learn the effects of covariate $X$ on the true binary outcome $Y^{*}$, but we only observe the reported outcome $Y$ which can be possibly misreported. We study the identifying power of two different instruments respectively: the first one is instrument $Z$ that only affects the true outcome $Y^{*}$ and the other is instrument $W$ that only affects the reported outcome $Y$ by affecting misreporting probabilities. Furthermore, we also study identification results of each instrument with additional assumptions on misreporting in Section \ref{subsec:add_mis}.


\subsubsection{Instrument $Z$}\label{subsec:z}

This section studies the identifying power of instrument $Z$ that only affects the true choice probability but does not influence misreporting probabilities. Instrument $Z$ affects the true variable $Y^*$ directly so it is a component of the covariate vector $X$. Therefore, we divide covariate $X$ into two parts: instrument $Z$ and the remaining covariates denoted as $\Xz=X\setminus Z\in \tilde{\mathcal{X}  }$.


Next we state some assumptions on instrument $Z$. 

\begin{ass}[Exclusion] \label{exc1}
For any $x\in \mathcal{X}, y\in\{0,1\}$, 
\begin{equation*}
\Pr(Y=1-y\mid Y^{*}=y, x)=\Pr(Y=1-y\mid Y^{*}=y, \xz).
\end{equation*}
\end{ass}
The exclusion restriction requires instrument $Z$ to be independent of misreporting process, so it only affects the reported probability by shifting the true probability. When the true outcome is participation in social programs, such as a job training program, one example of instrument $Z$ could be randomly assigned eligibility for the program. The random assignment will affect true participation but not influence misreporting, given its random nature. Studies, e.g., \cite{mahajan2006}, \cite{ura2018}, and \cite{ditraglia2019}, provide more examples of this instrument in various applications. 

 
The next assumption is about the extent of misreporting probabilities.

\begin{ass}[Degree of Misreporting] \label{deg1}
For any $\xz\in \tilde{\mathcal{X}}$, 
\begin{equation*}
\Pr(Y=0\mid Y^{*}=1, \xz)+\Pr(Y=1\mid Y^{*}=0, \xz) \leq 1.
\end{equation*}
\end{ass}

Assumption \ref{deg1} is about the degree of misreporting, ensuring that the reported data is informative for the true probability. It requires that the misreporting errors are not too large so that the sum of two-sided misreporting probabilities is smaller than one. This assumption is consistent with empirical evidence in multiple studies such as \cite*{meyer2009} and \cite*{meyer2020}, which document that one-sided misreporting probability is usually less than 50\% in survey data. Moreover, our identification analysis only requires the degree of misreporting to be known, and the analysis is applicable when the sum of two-sided misreporting rates is larger than one.

\begin{ass}[Boundary Condition]\label{bou1}
The reported choice probability $p(x)$ satisfies that: $\sup \limits_{z\in \mathcal{Z}}p(\xz, z)>0$ and $\inf\limits_{z\in \mathcal{Z}}p(\xz, z)<1$ for any $\tilde{x}\in \tilde{\mathcal{X} }$.
\end{ass}
Assumption \ref{bou1} is a boundary condition for the reported probability. This assumption is relatively weak, requiring that the supremum of the reported probability is bounded away from zero and the infimum of the reported probability is bounded away from one. It can be satisfied when instrument $Z$ strictly affects the reported probability and takes on at least two values.

Under the above assumptions, we are ready to establish identification results for the true probability $p^{*}(x)$.

\begin{prop}\label{prop_z}
Under Assumptions \ref{exc1}-\ref{bou1}, the sharp bounds for $p^{*}(x)$ are characterized as $p^{*}(x)\in [L_{1}(x), U_{1}(x) ]$ for any $x\in \mathcal{X}$, where
\begin{equation*}
L_{1}(x)=\frac{p(x)- \under{p}_z(\xz)}{1- \under{p}_z(\xz)},  \qquad   U_{1}(x)=\frac{p(x)}{\bar{p}_z(\xz)}.
\end{equation*}
where $\under{p}_z(\xz):= \inf\limits_{z\in \mathcal{Z}} p(\xz, z) $ and $\bar{p}_z(\xz):=\sup \limits_{z\in \mathcal{Z}} p(\xz, z)$.

\end{prop}

Proposition \ref{prop_z} characterizes partial identification for the true probability $p^{*}(x)$ using variation in instrument $Z$. It also establishes sharpness of the results, showings that the bounds are the best possible given the assumptions and data. 
According to the definition of the bounds, we know that $L_1(x), U_1(x)\in [0, 1]$. The identifying power of instrument $Z$ depends on the range of the reported probability $p(\tilde{x}, z)$ as one varies $z$: tighter bounds for the true probability are achieved with a larger range of the reported probability. A larger variation in the reported probability can yield smaller bounds for misreporting probabilities, leading to tighter bounds on the true probability.

When the conditional reported probability can vary from zero to one when instrument $Z$ changes, we can infer that there is no misreporting and achieve point identification as $p^{*}(x)=p(x)$. The intuition is as follows. When the reported probability is one (or zero), there are two possibilities: one is that the true conditional probability is one, and individuals with $Y^{*}=1$ all report the truth (or misreport); the other is that the true probability is zero while  individuals with $Y^{*}=0$ all misreport (or report the truth). 
The possibility that everyone misreports can be rejected by Assumption \ref{deg1}, which requires that the sum of two-sided misreporting probabilities to be smaller than one. Therefore, we can conclude that there is no misreporting, and point identification for the true probability is obtained.

\subsubsection{Instrument $W$}\label{subsec:w}

This section introduces an alternative method for identifying the true probability using instrument $W$. This instrument is assumed to only affect misreporting probabilities but not the true probability. In addition to covariate $X$ in the true binary choice model, instrument $W$ is an additional variable related to misreporting processes but excluded from the true binary choice model. Our approach does not impose any parametric models for misreporting processes, allowing instrument $W$ to affect misreporting probabilities nonparametrically.

The complete set of observed variables are $(X, W, Y)$ in this section. Let $\p(x, w)=\Pr(Y=1\mid x, w)$ denote the reported probability conditional on $(X, W)=(x, w)$. The subscript $W$ in the function $\p$ is used to distinguish it from the function $p(x)=\Pr(Y=1\mid x)$.

The following presents assumptions on instrument $W\in \mathcal{W}$.

\begin{ass} [Exclusion] \label{exc2}
\begin{equation*}
\Pr(Y^{*}=1\mid x, w)=\Pr(Y^{*}=1\mid x)=p^{*}(x),        
\end{equation*}
for any $ x\in \mathcal{X}$ and $w\in\mathcal{W}$.  
\end{ass}

Similar to Assumption \ref{exc1} for instrument $Z$, Assumption \ref{exc2} states a different exclusion restriction, requiring instrument $W$ to be independent of the true dependent variable $Y^*$. Examples for instrument $W$ could include interview-related variables, such as interviewersÕ assessments of respondentsÕ accuracy or different interview styles like phone interviews and in-person interviews. These variables are unlikely to affect the true dependent variable but are related to respondentsÕ probabilities of reporting the truth. These variables are unlikely to affect the true dependent variable, but are relevant to respondents' probabilities of reporting the truth.


\begin{ass}[Monotonicity] \label{mono}
For any $x\in\mathcal{X}$, the following conditions hold for any $y\in\{0, 1\}$ and $w_1>w_2\in \mathcal{W}$,
\begin{equation*}
\begin{aligned}
\Pr(Y=1-y\mid Y^{*}=y, x, w_1) \leq \Pr(Y=1-y\mid Y^{*}=y, x, w_2).  \\
\end{aligned}
\end{equation*}
\end{ass}

Assumption \ref{mono} is the monotonicity condition for instrument $W$: the misreporting probabilities are weakly decreasing with respect to the instrument. For instance, when the instrument is interviewers' evaluations of respondents' accuracy, it is natural that misreporting probabilities are smaller with higher evaluations. In the case of interview styles, individuals are more likely to report the truth during an in-person interview than a phone interview. It is also worth noting that Assumption \ref{mono} is a weak monotonicity condition, as it only requires monotonicity for the average probability but allows for potential violations in certain populations.

\begin{ass} \label{bou2}

\begin{enumerate}[(1)]
\item Degree of Misreporting: for any $x\in \mathcal{X}, w\in \mathcal{W}$,
\begin{equation*}
\Pr(Y=1\mid Y^{*}=0, x, w)+\Pr(Y=0\mid Y^{*}=1, x, w)\leq 1.
\end{equation*}

\item Boundary Condition: the reported probability $\p(x, w)$ is bounded away from zero and one: $0<\p(x, w)<1$ for any $x\in \mathcal{X}, w\in \mathcal{W}$.
\end{enumerate}
\end{ass}
Assumption \ref{bou2}  is similar to Assumptions \ref{deg1}-\ref{bou1} for instrument $Z$. The difference is that all probabilities are also conditional on the additional variable $W$ since the misreporting probabilities and the reported probability depend on instrument $W$ in this scenario. 

Under the above assumptions, the next proposition establishes identification results for the true probability $p^{*}(x)$.
\begin{prop}\label{prop_w}
Under Assumptions \ref{exc2}-\ref{bou2}, the true probability $p^{*}(x)$ can be bounded as $p^{*}(x)\in [L_2(x), U_2(x)]$ for any $x\in \mathcal{X}$, where
\begin{equation*}
L_2(x)= \sup_{w\in \mathcal{W}} \left\{  \frac{\p(x, w)- \under{p}_w(x, w)}{1- \under{p}_w(x, w) } \right\},
 \quad
 U_2(x)= \inf_{w\in \mathcal{W}} \left\{ \frac{\p(x, w)}{ \bar{p}_w(x, w) } \right\},
\end{equation*}
where $\under{p}_w(x, w):= \inf\limits_{\tilde{w} \leq w } \p(x, \tilde{w})$ and $\bar{p}_w(x, w):= \sup\limits_{\tilde{w}\leq w  } \p(x, \tilde{w})$. Moreover, the above bounds are sharp when instrument $W$ is binary.
\end{prop}

Proposition \ref{prop_w} derives partial identification by using a distinct instrument that only monotonically influences misreporting probabilities but not the true probability. Furthermore, it demonstrates that this result exhausts all possible information when the instrument $W$ is binary. The bounds, as per their definitions, satisfy that $L_2(x), U_2(x)\in[0, 1]$. Additionally, the results in Proposition \ref{prop_w} imply that $U_2(x)\geq L_2(x)$ for all $x$, providing testable implications for our assumptions. 

Proposition \ref{prop_w} mainly exploits the exclusion and monotonicity of instrument $W$. Under the monotonicity condition, the misreporting probabilities at $w$ can be bounded above by all upper bounds of misreporting probabilities evaluated at smaller values $\tilde{w}\leq w$. The identifying power of monotonicity is shown within the bracket of the bounds $(L_2(x), U_2(x))$.
The exclusion restriction can help tighten the bounds for the true probability $p^{*}(x)$ by intersecting all bounds derived from any value of instrument $W$, as demonstrated outside the bracket of bounds.

\subsection{Additional Restrictions on Misreporting} \label{subsec:add_mis}

Our analysis can be also combined with other additional information on misreporting probabilities to further tighten the bounds for the true probabilities $p^{*}(x)$. Section \ref{subsec:oneside}-\ref{subsec:mono_mis} studies one-sided misreporting, bounded misreporting probabilities, as well as monotone misreporting probabilities.

\subsubsection{One-sided Misreporting} \label{subsec:oneside}

This section studies identification of the true probability $p^{*}(x)$ under one-sided misreporting. One-sided misreporting refers to where only one group with the true outcome $Y^{*}=y$ misreport, while the other group with $Y^{*}=1-y$ always report the truth. This assumption has practical applications and has been used in previous studies, such as \cite*{nguimkeu2019}. 
In their paper, they investigate the participation in the food stamp program and provide evidence for a small overreporting probability. The assumption regarding which group has no misreporting depends on the specific application, and we present results for both cases.

The next proposition provides bounds for the true probability $p^{*}(x)$ under one-sided misreporting by using one of the two instruments $(Z, W)$ respectively.

\begin{prop} \label{prop_one}

(1) Under Assumptions \ref{exc1}-\ref{bou1}, the sharp bounds for the true probability $p^{*}(x)$ are characterized as
\begin{equation*}
p^{*}(x)\in \left\{
\begin{aligned}
&\left[p(x),  U_1(x) \right]  \qquad \text{when} \ \Pr(Y=1\mid Y^{*}=0,\xz)=0, \\
&\left[L_1(x),  p(x) \right]  \qquad \text{when}  \ \Pr(Y=0\mid Y^{*}=1,\xz)=0.
\end{aligned}
\right.
\end{equation*}

(2) Under Assumptions \ref{exc2}-\ref{bou2}, the sharp bounds for the true probability $p^{*}(x)$ are characterized as
\begin{equation*}
p^{*}(x)\in \left\{
\begin{aligned}
&\left[\sup_{w\in \mathcal{W}} \p(x, w),  U_2(x) \right]  \qquad \text{when} \ \Pr(Y=1\mid Y^{*}=0, x, w)=0, \\
&\left[L_2(x),  \inf_{w\in \mathcal{W}} \p(x, w) \right]  \qquad \text{when}  \ \Pr(Y=0\mid Y^{*}=1, x, w)=0. 
\end{aligned}
\right.
\end{equation*}

\end{prop}

Proposition \ref{prop_one} provides sharp identification results for $p^{*}(x)$ under different scenarios of one-sided misreporting, using one of the two instruments, respectively. The one-sided misreporting assumption ensures accurate data from one group and establishes the misreporting probability to be zero for this group. In comparison to the results in Proposition \ref{prop_z} and \ref{prop_w}, which allow for two-sided misreporting, Proposition \ref{prop_one} consistently provides tighter bounds on $p^{*}(x)$ with either larger lower bound or smaller upper bound.\footnote{When instrument $W$ is available, the one-sided misreporting assumption implies the monotonicity of $\p(x, w)$ in $w$, and the upper bound $U_2(x)$ becomes one. This implication can serve as a testable implication for the one-sided misreporting assumption. }


\subsubsection{Bounded Misreporting Probabilities} \label{subsec:bound_mis}

This section considers that the misreporting probabilities are bounded by known numbers. This restriction may come from information such as previous studies on misreporting, auxiliary administrative data, or theoretical models that provide insights into the potential range of misreporting probabilities.

\begin{ass} \label{ass:bounds_mis}
The misreporting probabilities satisfy the following condition: for $y \in \{ 1, 2\}$ and any $(x, w)$,
\[
\Pr(Y=1-y \mid Y^{*}=y, x, w) \leq \bar{\a}_y, 
\]
where $\bar{\a}_y \in [0, 1]$ is a known constant.
\end{ass}

Assumption \ref{ass:bounds_mis} may stem from empirical results in other research or auxiliary information, such as administrative data from other samples. For simplicity, Assumption \ref{ass:bounds_mis} adopts uniform bounds for misreporting probabilities regardless of values of $(x, w)$. Our approach can be also adjusted to allow $\bar{\a}_y$ to depend on $(x, w)$.  When instrument $W$ is not available, then the above conditional misreporting probability can be adjusted by only conditioning on the covariate $x$.

\begin{prop} \label{prop:bounds_mis}

(1) Under Assumptions \ref{exc1}-\ref{bou1} $\&$ \ref{ass:bounds_mis}, the true probability $p^{*}(x)$ can be bounded as
\begin{equation*}
p^{*}(x)\in  \left[\frac{p(x)- \min \left\{ \under{p}(\xz), \bar{\a}_0 \right\}}{1-  \min \left\{ \under{p}(\xz), \bar{\a}_0 \right\} },  \frac{p(x)}{\max \left\{ \bar{p}(\xz), 1-\bar{\a}_1 \right\} } \right]; 
\end{equation*}
(2) Under Assumptions \ref{exc2}-\ref{bou2} $\&$ \ref{ass:bounds_mis}, the true probability $p^{*}(x)$ can be bounded as
\begin{equation*}
\begin{aligned}
p^{*}(x)\in \left[ \sup_{w\in \mathcal{W}} \left\{  \frac{\p(x, w)- \min \left\{   \under{p}_w(x, w), \bar{\a}_0  \right\} }{1- \min \left\{  \under{p}_w(x, w), \bar{\a}_0 \right\} } \right\},  \inf_{w\in \mathcal{W}} \left\{   \frac{\p(x, w)}{\max\left \{ \bar{p}_w(x, w),  1-\bar{\a}_1 \right\}  }   \right\} \right].
\end{aligned}
\end{equation*}
\end{prop}

Proposition \ref{prop:bounds_mis} derives partial identification for $p^{*}(x)$through additional bounds on misreporting probabilities, with the identifying power depending on the value of the bounds $\bar{\a}_y$. A smaller value of $\bar{\a}_y$ results in tighter bounds for $p^{*}(x)$, and $\bar{\a}_y=0$ corresponds to no misreporting. It is also possible that this additional restriction may not provide any information if the value of the bound $\bar{\a}_y$ is larger than that derived by using each instrument.

\subsubsection{Monotone Misreporting Probabilities} \label{subsec:mono_mis}

This section explores monotone misreporting probabilities,  where the misreporting probability of one group is smaller than that of the other group. The direction of monotonicity depends on 
specific applications. To illustrate, we study the following monotonicity on misreporting probabilities.

\begin{ass} \label{ass:mono_mis}
The misreporting probabilities satisfy the following condition: for any $(x, w)$,
\[
\Pr(Y=1 \mid Y^{*}=0, x, w) \leq \Pr(Y=0 \mid Y^*=1, x, w).
\]
\end{ass}

Assumption \ref{ass:mono_mis} says that the misreporting probability for people with $Y^*=0$ is smaller than those with $Y^*=1$. This assumption is applicable for scenarios where $Y^*$ represents participation in social assistance programs or whether one is smoking or not. It is documented in the literature that people who did not participate in social program are more likely to report the truth than people who participated. The direction of the monotonicity can be reversed in some applications, such as $Y^*$ represents educational attainment. 
Our identification methods can accommodate both cases, where the direction of the monotonicity is required to be known.

\begin{prop} \label{prop:mono_mis}

(1) Under Assumptions \ref{exc1}-\ref{bou1} $\&$ \ref{ass:mono_mis}, the true probability $p^{*}(x)$ can be bounded as
\begin{equation*}
p^{*}(x)\in  \left[\frac{p(x)-  \min\left\{ \under{p}_z(\xz), 1-\bar{p}_z(\xz) \right\} }{1-  \min\left\{ \under{p}_z(\xz), 1-\bar{p}_z(\xz) \right\} }, \frac{p(x)}{ \bar{p}(\xz) } \right]; 
\end{equation*}
(2) Under Assumptions \ref{exc2}-\ref{bou2} $\&$ \ref{ass:mono_mis}, the true probability $p^{*}(x)$ can be bounded as
\begin{equation*}
\begin{aligned}
p^{*}(x)\in \left[ \sup_{w\in \mathcal{W}} \left\{  \frac{\p(x, w)- \min \left\{   \under{p}_w(x, w), 1-\bar{p}_w(x, w)  \right\} }{1- \min \left\{  \under{p}_w(x, w), 1-\bar{p}_w(x, w) \right\} } \right\},  \inf_{w\in \mathcal{W}} \left\{   \frac{\p(x, w)}{ \bar{p}_w(x, w) }   \right\} \right].
\end{aligned}
\end{equation*}
\end{prop}

Proposition \ref{prop:mono_mis} shows that  Assumption \ref{ass:mono_mis} can further increase the lower bound for $p^{*}(x)$ with each instrument. The idea is that under the monotone misreporting rates assumption, the misreporting probability for group $Y^*=0$ can also be bounded by the upper bound of the misreporting probability for the other group $Y^*=1$, yielding more informative results for the true probability $p^*(x)$. Symmetrically, when assuming a smaller misreporting rate for group $Y^*=1$ compared to $Y^*=0$, the upper bound for $p^{*}(x)$ can be tightened.

~

\section{Conditional Moment Inequalities}\label{sec:cmi}

Based on previous identification analysis, this section demonstrates how to characterize identification for various binary choice models using conditional moment inequalities. We focus on the two binary choice models with cross-sectional data in Examples \ref{example:par} and \ref{example:semi} to illustrate the idea, and the analysis can be applied to panel binary choice models.\footnote{For panel models, we can follow the same identification strategy to derive bounds for the true probability $p^{*}_t(x):=E[Y_t^*\mid X_{st}= x]$ at each period $t$. The main distinction is that all variables and results will be indexed by time period $t$.}

\subsection{Parametric Binary Choice Model}

As shown in Example \ref{example:par}, given the parametric structure of the error term $\e$, i.e., $\e\mid X \sim F_{\e}(\cdot)$, the model parameter $\b_0$ is characterized by the following restriction: 
\begin{equation} \label{equ:par}
L_k(x) \leq F_{\e}(x'\b_0) \leq U_k(x),
\end{equation}
where $k \in\{1, 2\}$ depends on the availability of instruments.

When the instrument $Z$ is available ($k=1$), plugging into the definition of $(L_1(x), U_1(x))$,  the identifying restriction in \eqref{equ:par} is equivalent to the following conditional moment inequality: 
\[E[g_{z, par}(Y, X, \b_0)\mid x]\geq 0,
\]
where $g_{z, par}(Y, X, \b_0))$ is defined as
\[g_{z, par}(Y, X, \b_0)):=\left\{\begin{aligned}
&Y-F_{\e}(X'\b_0) \bar{p}_Z(\Xz) \\
& F_{\e}(X'\b_0)  + \under{p}_Z(\Xz)  (1-F_{\e}(X'\b_0)  ) -Y.
\end{aligned}
\right.
\]

Similarly, when the instrument $W$ is available  ($k=2$), condition \eqref{equ:par} is equivalent to the restriction: 
\[
E[g_{w, par}(Y, X, W, \b_0)\mid x, w]\geq 0,\]
 where $g_{w, par}(Y, X, W, \b_0)$ is defined as
\[g_{w, par}(Y, X, W, \b_0):=\left\{\begin{aligned}
&Y-F_{\e}(X'\b_0) \bar{p}_W(X, W) \\
& F_{\e}(X'\b_0)  + \under{p}_W(X, W)  (1-F_{\e}(X'\b_0)  ) -Y.
\end{aligned}
\right.
\]

\subsection{Semiparametric Binary Choice Model}

Under the semiparametric framework in Example \ref{example:semi}, the identifying restriction of the model parameter $\b_0$ is characterized as
\begin{equation}\label{equ:semi}
\begin{aligned}  
x'\b_0 \geq 0\Longrightarrow  U_k(x)-0.5\geq 0, \\
x'\b_0 \leq 0 \Longrightarrow  L_k(x)-0.5\leq 0,
\end{aligned}
\end{equation}
for $k \in \{1, 2\}$. To characterize conditional moment inequalities, we first conduct monotone transformations of the bounds $(L_k(x), U_k(x))$ by multiplying them by their respective (positive) denominators. This monotone transformation preserves the sign of those bounds so it does not affect the identification results. 

When instrument $Z$ is available with $k=1$, the following condition holds by multiplying the denominators of $(L_1(x), U_1(x))$:
\begin{equation*}
\begin{aligned}
U_1(x)-0.5\geq 0 & \Longleftrightarrow p(x)-0.5 \bar{p}_Z(\xz)\geq 0, \\
L_1(x)-0.5\leq 0 &\Longleftrightarrow  p(x)-0.5 \under{p}_Z(\xz) -0.5\leq 0. \\
\end{aligned}
\end{equation*}


Then, restriction  \eqref{equ:semi} is equivalent to the following conditional moment restriction:
\[
E[g_{z, semi}(Y, X, \b_0)\mid x]\geq 0,
\]
where $g_{z, semi}(Y, X, \b_0)$ is defined as
\[g_{z, semi}(Y, X, \b_0):=\left\{\begin{aligned}
&X'\beta_{0}\mathbbm{1}\{X'\beta_{0}\geq 0\}(Y-0.5 \bar{p}_Z(\Xz)), \\
& X'\beta_{0}\mathbbm{1}\{X'\beta_{0} \leq 0 \}(Y-0.5 \under{p}_Z(\Xz)-0.5).
\end{aligned}
\right.
\]

When the instrument $W$ is available with $k=2$, we can conduct a similar monotone transformation of $(L_2(x), U_2(x))$, yielding the following relationship:
\begin{equation*}
\begin{aligned}
&U_2(x)-0.5\geq 0\Longleftrightarrow p_W(x, w)-0.5 \bar{p}_W(x, w) \geq 0 \quad \forall w, \\
&L_2(x)-0.5\leq 0 \Longleftrightarrow   p_W(x, w)-0.5\under{p}_W(x, w)-0.5 \leq 0  \quad \forall w.
\end{aligned}
\end{equation*}

Based on the above relationship, restriction  \eqref{equ:semi} is equivalent to the following conditional moment restriction:
\[
E[g_{w, semi}(Y, X, W, \b_0)\mid x, w]\geq 0,
\]
where $g_{w, semi}(Y, X, W, \b_0)$ is defined as
\[g_{w, semi}(Y, X, W, \b_0):=\left\{\begin{aligned}
&X'\beta_{0}\mathbbm{1}\{X'\beta_{0}\geq 0\}(Y - 0.5 \bar{p}_W(X, W)), \\
& X'\beta_{0}\mathbbm{1}\{X'\beta_{0} \leq 0 \}(Y - 0.5 \under{p}_W(X, W)-0.5).
\end{aligned}
\right.
\]

~

For both parametric and semiparametric models, we characterize partial identification of the model parameter using conditional moment inequalities. Then, we can adopt established methods from the literature developed for general conditional moment inequalities to conduct estimation and inference, such as  \cite*{chernozhukov2007}, \cite{andrews2013}, and \cite*{chernozhukov2013}.\footnote{The conditional moment inequalities contain nuisance parameters, e.g., $\bar{p}_Z(\xz)$ and $\under{p}_Z(\xz)$, that can be consistently estimated. The inference methods in the literature, such as \cite{andrews2013} Section 8,  allow for preliminary consistent estimation of nuisance parameters.}

\section{Simulation Study} \label{sec:sim}

This section examines the finite sample performance of our identification approaches via Monte Carlo simulations. We focus on the semiparametric binary choice model presented in Example \ref{example:semi}, which allows for flexible misreporting process and does not require distributional assumption. To better evaluate our approach, we also implement the maximum likelihood estimation approach proposed in  \cite*{hausman1998} for comparison, referred to as the HAS approach. Their method accounts for potential misreporting, but assumes distributional assumption on error term and homogeneous (constant) misreporting probabilities. 
The simulation results demonstrate the robustness of our approach concerning heterogeneous misreporting probabilities and parametric assumptions.

For our approach, we follow \cite*{chernozhukov2007} and \cite{andrews2013} to estimate the identified set based on conditional moment inequalities. We first transform conditional moment inequalities into unconditional moment inequalities using  indicator functions of hypercubes in the space of covariates as instrumental functions.\footnote{See \cite{andrews2013} for more choices and discussions of instrumental functions.} The number of hypercubes is $\{30, 40, 50\}$ for the sample size $n\in \{1000, 2000, 4000\}$.  Then the estimated identified set can be computed based on the criterion function method in \cite*{chernozhukov2007}.

Next, we present the performance of our approach and the HAS method in \cite*{hausman1998} using different instruments.

\subsection{Instrument $Z$}

This section studies the identification results using instrument $Z$.
The DGP is described as follows. Instrument $Z$ is uniformly distributed over the set  $\{-1, -0.5, 0, 0.5, 1\}$, $\tilde{X}$ follow a uniform distribution over the interval $[-1, 1]$, and the full covariate $X$ is given as $X=[1; \tilde{X}; Z]$. The true parameter $\beta_0=[1; 1.5; -1.5]$ and the true outcome is generated by $Y^{*}= \mathbbm{1}\left\{X'\beta_0 \geq \epsilon \right\}$. We study two specifications of the error term $\epsilon$: a standard normal distribution $\mathcal{N}(0, 1)$ and a Cauchy distribution $Cauchy(0, 0.5)$. 

The reported outcome $Y$ is given by $Y =M_{1}\cdot Y^*+(1-M_{0})\cdot (1-Y^*)$, where $M_y\in \{0, 1\}$ denotes the reporting variable and $M_y=0$ represents misreporting for any $y\in \{0, 1\}$. We allow for heterogeneous misreporting probabilities, depending on the value of covariate $\tilde{X}$:
\[
\Pr(M_1=0\mid \tilde{x})=0.1 - 0.1 \tilde{x}, \quad \Pr(M_0=0\mid \tilde{x})=0.3 + 0.1 \tilde{x}.
\]

In this DGP, it is clear that  Assumptions \ref{exc1} and \ref{deg1} on the exclusion and degrees of misreporting are satisfied. The sample size is $n \in \{500, 1000, 2000\}$ and repetition number is $B = 300$.

To compare the performance of the two methods, we report the root mean-squared error (rMSE) and median of absolute deviation (MAD) for the lower bound $\hat{\beta}^{l}$ and upper bound $\hat{\beta}^u$ in this paper, along with the parametric estimator $
\hat{\beta}_{HAS}$, which assumes constant misreporting probabilities and a normal distribution of the error term.  Let $\beta^k$ denote the $k$th element of the parameter $\beta$. We normalize the first element of the parameter to one for our method: $\b_0^1=1$.

\begin{table}[!htbp]
\centering
\caption{Performance Comparisons for $\hat{\beta}^2$: Instrument $Z$}
\label{table:z1}
\begin{tabular}{c|cccc|cc}
\hline
\hline
 \rule{0pt}{20pt}\multirow{2}{*}{Design}& \multicolumn{2}{c}{$\hat{\beta}_{semi}^l$} & \multicolumn{2}{c|}{$\hat{\beta}_{semi}^u$}    &  \multicolumn{2}{c}{$\hat{\beta}_{HAS}$}  \\
\cline{2-7} 
\rule{0pt}{20pt}  & rMSE & MAD& rMSE & MAD & rMSE & MAD \\
 \hline
 \rule{0pt}{20pt}&    \multicolumn{6}{c} {$n=500$} \\
\hline
 \rule{0pt}{20pt} Normal &  0.482  &0.338  &0.468  &0.338 & 0.475  &0.265 \\
 \rule{0pt}{20pt} Cauchy &   0.518 &  0.338  & 0.500 &  0.338 &  1.115  & 0.407 \\
 \hline
\rule{0pt}{20pt}&  \multicolumn{6}{c} {$n = 1000$} \\
\hline
\rule{0pt}{20pt}Normal &  0.434 & 0.338 & 0.408 & 0.268&  0.302 & 0.245  \\                                                                                                                                                                                                                                                                                                                                                                                                                                                                                                                                                                                                                                                              
\rule{0pt}{20pt}Cauchy  & 0.385  & 0.268  & 0.356 &  0.237 &  0.703  & 0.349    \\ 
\hline
\rule{0pt}{20pt} & \multicolumn{6}{c} {$n = 2000$} \\
 \hline
 \rule{0pt}{20pt}Normal & 0.363  & 0.338   &0.331   &0.237  & 0.242  & 0.180 \\                                                                                                                                                                                                                                                                                                                                                                                                                                                                                                                                                                                                                                                                      
\rule{0pt}{20pt}Cauchy  & 0.376 & 0.268&  0.323 &  0.237 & 0.438 & 0.274 \\ 
\hline
\end{tabular}
\end{table}

\begin{table}[!htbp]
\centering
\caption{Performance Comparisons for $\hat{\beta}^3$: Instrument $Z$}
\label{table:z2}
\begin{tabular}{c|cccc|cc}
\hline
\hline
 \rule{0pt}{20pt}\multirow{2}{*}{Design}& \multicolumn{2}{c}{$\hat{\beta}_{semi}^l$} & \multicolumn{2}{c|}{$\hat{\beta}_{semi}^u$}    &  \multicolumn{2}{c}{$\hat{\beta}_{HAS}$} \\
\cline{2-7} 
\rule{0pt}{20pt} & rMSE & MAD & rMSE & MAD  & rMSE & MAD  \\
 \hline
 \rule{0pt}{20pt}&    \multicolumn{6}{c} {$n=500$} \\
\hline
 \rule{0pt}{20pt} Normal &  0.519 & 0.369  &0.527 & 0.439 & 0.587 & 0.563  \\
 \rule{0pt}{20pt} Cauchy & 0.544 &  0.439 &  0.566  & 0.439 &   0.899 &  0.573 \\
 \hline
\rule{0pt}{20pt} & \multicolumn{6}{c} {$n =1000$} \\
 \hline
 \rule{0pt}{20pt}Normal & 0.427 & 0.338 & 0.449 & 0.338 & 0.602 & 0.617  \\                                                                                                                                                                                                                                                                                                                                                                                                                                                                                                                                                                                                                                                                      
\rule{0pt}{20pt}Cauchy  & 0.452 &  0.338&   0.485&   0.439&   0.640 &  0.472  \\ 
\hline
\rule{0pt}{20pt} & \multicolumn{6}{c} {$n = 2000$} \\
 \hline
\rule{0pt}{20pt}Normal &  0.376  & 0.237  & 0.391   &0.338 &  0.596 &  0.583  \\
\rule{0pt}{20pt}Cauchy & 0.351 & 0.237&  0.393 & 0.338 & 0.479 & 0.416  \\
\hline
\end{tabular}
\end{table}

Tables \ref{table:z1} and \ref{table:z2} display the performance of the two methods under different specifications of the error term $\epsilon$ and different sample sizes. The results illustrate that our approach uniformly performs well across various error term specifications.  When the distribution of $\e$ is correctly specified, the HAS method shows reasonable performance but does not necessarily outperform our method, as the constant misreporting probabilities assumption still remains misspecified. Moreover,  the HAS estimator exhibits significant bias when the distributional assumption is also misspecified (under the Cauchy design), while our approach has robust performance under different designs.

\subsection{Instrument $W$}

This section examines the finite sample performance of the identified set using instrument $W$. The DGP is described as follows. Covariate $X$ is given by $X=(1; \tilde{X})$, where $\tilde{X}$ follows a uniform distribution over $[-1,1]$. The true outcome $Y^{*}$ is generated by $Y^{*}=\mathbbm{1}\{\epsilon\leq X'\beta_0 \}$, where the true parameter is $\beta_0=[1; 1.5]$. Similarly, we consider two specifications of error term $\epsilon$: a standard normal distribution $\mathcal{N}(0, 1)$ and a Cauchy distribution $Cauchy(0, 0.5)$. 

Instrument $W$ is uniformly distributed over the set $\{1, 2, 3, 4, 5\}$. The reported outcome $Y$ is given by $Y =M_{1}\cdot Y^*+(1-M_{0})\cdot (1-Y^*)$, where $M_y\in \{0, 1\}$ and $M_y=0$ denotes misreporting. The misreporting probability $\Pr(M_y =0\mid \tilde{x}, w)$ depends on covariate $\tilde{X}$ and instrument $W$ as follows:
\[
\Pr(M_1=0\mid \tilde{x}, w)=0.1 - 0.1 \tilde{x}, \quad \Pr(M_0=0\mid \tilde{x}, w)=\frac{1}{1+0.3w^2}.
\]


In this specification, it can be verified that the monotonicity condition in Assumption \ref{mono} on instrument $W$ and the degree of misreporting probabilities in Assumption \ref{bou2} are satisfied. The sample size is $n \in \{500, 1000, 2000\}$ and repetition number is $B = 300$.

\begin{table}[!htbp]
\centering
\caption{Performance Comparisons for $\hat{\beta}^2$: Instrument $W$}
\label{table:w}
\begin{tabular}{c|cccc|cc}
\hline
\hline
 \rule{0pt}{20pt}\multirow{2}{*}{Design}& \multicolumn{2}{c}{$\hat{\beta}_{semi}^l$} & \multicolumn{2}{c|}{$\hat{\beta}_{semi}^u$}    &  \multicolumn{2}{c}{$\hat{\beta}_{HAS}$} \\
\cline{2-7} 
\rule{0pt}{20pt} & rMSE & MAD & rMSE & MAD  & rMSE & MAD \\
 \hline
 \rule{0pt}{20pt}&    \multicolumn{6}{c} {$n = 500$} \\
\hline
 \rule{0pt}{20pt} Normal & 0.567 & 0.338  &0.561 &  0.338  &2.342 & 1.735\\
\rule{0pt}{20pt} Cauchy &   0.598 &  0.439 &  0.587 &  0.338 &  2.966  & 3.500   \\
 \hline
\rule{0pt}{20pt}&  \multicolumn{6}{c} {$n = 1000$} \\
\hline
\rule{0pt}{20pt} Normal &   0.423 & 0.237 & 0.412 & 0.237 & 2.244 & 1.456     \\        
\rule{0pt}{20pt} Cauchy  &    0.391 &  0.237 &  0.372  & 0.237  & 2.739  & 3.081  \\                                                                                                                                                                                                                                                                                                                                                                                                                                                                                                                                                                                                                                                                   
\hline
\rule{0pt}{20pt} & \multicolumn{6}{c} {$n = 2000$} \\
 \hline
 \rule{0pt}{20pt} Normal &  0.363 & 0.237 & 0.351  &0.167 & 1.935 & 1.128    \\      
\rule{0pt}{20pt} Cauchy  &  0.341 & 0.237 &  0.334 & 0.237 &  2.736 & 3.098   \\                                                                                                                                                                                                                                                                                                                                                                                                                                                                                                                                                                                                                                                                     
\hline
\end{tabular}
\end{table}

Table \ref{table:w} displays the performance of $\beta^2_0$ under different sample sizes and specifications of $\epsilon$. The results show that our approach consistently outperforms the HAS approach across all specifications. The HAS method exhibits a significant bias even when the distribution of  $\e$ is correctly specified, as the range of misreporting probabilities (under different values of $(\tilde{x}, w)$) is very large in this DGP and the constant misreporting probabilities assumption is seriously misspecified.
The bias of HAS method becomes larger when the distributional assumption is also misspecified under the Cauchy design.
In summary, the simulation results demonstrate the robust performance of our two approaches concerning flexible misreporting processes and distributional assumptions.

\section{Empirical Illustration} \label{sec:emp}

As an empirical illustration, we apply our methods to analyze educational attainment using a binary choice model with potential misreporting. The dataset we use is drawn from the National Longitudinal Surveys in 1976 (NLSY76), which is also used in \cite{card1995} to estimate returns to education. This survey data contains 3613 individuals' self-reported information including educational experiences and family backgrounds. The objective is to explore how people's characteristics affect the probability of them attaining a college degree. However, there may be misreporting in self-reports of educational attainment in this data, which could severely bias the estimation results.

In this application, the reported outcome $Y$ is whether an individual reports attending a college which may be subject to misreporting. Instrument $Z$ is whether an individual grew up near a four-year college (college proximity). This instrument affects people's true decision of attending college, but may not affect their misreporting behaviors. We also include two other covariates in the binary choice model: parents' average education $X_1$ and whether an individual is black $X_2$. The following table shows the summary statistics of all variables.

\begin{table}[!htbp]
\centering
\caption{Summary Statistics}
\label{table:summary}
\setlength{\tabcolsep}{15pt}
\renewcommand{\arraystretch}{1.5}
\captionsetup[table]{skip=10pt}
\begin{tabular}{c|cccc}
\hline
\hline
& $Y$ & $X_1$  & $X_2$ & $Z$ \\
\hline
min & 0  & 0  &  0 & 0\\
\hline
max & 1  & 18  & 1 & 1 \\
\hline
mean & 0.268 & 10.173 & 0.230 & 0.678 \\
\hline
s.d. & 0.443  & 2.786 & 0.421 & 0.467 \\
\hline
\end{tabular}
\end{table}

We adopt the semiparametric binary choice model with instrument $Z$ to study how individuals' observed characteristics affect the likelihood of attending a college.\footnote{The interview-related information is only available in a ``restricted-use" version of NLSY79, so there is no instrument $W$ for this application.}  In addition to accommodating flexible misreporting processes, our method is robust to distributional assumptions.
The full vector of covariate is $X=[1; X_1; X_2; Z]$, and the corresponding coefficient is denoted as $\beta_0=[\beta_0^0, \beta_0^1, \beta_0^2, \beta_0^3]$. For the semiparametric binary choice model, the coefficient $\beta_0$ can be only identified up to a constant. The coefficient of instrument $Z$ is normalized to one since people tend to be more likely to attend a college when they live closer to a college. For comparison, we also display the results of the HAS approach, which is  described in Section \ref{sec:sim}.

\begin{table}[!htbp]
\centering
\caption{Application: Estimation Results}
\label{table:edu}
\setlength{\tabcolsep}{10pt}
\renewcommand{\arraystretch}{1.5}
\captionsetup[table]{skip=10pt}
\begin{tabular}{c|c|c|c|c}
\hline
\hline
 & $\hat{\beta}^{0}$ & $\hat{\beta}^1$ & $\hat{\beta}^2 $  & $\hat{\beta}^3$ \\ 
\hline
this paper  &[-1.667, -0.879] &  [0.030, 0.091] &  [-1, -0.333] & 1 \\ 
\hline
HAS  &-32.577 & -23.368 & -10.582   &  -27.823\\ 
\hline
\end{tabular}
\end{table}

Table \ref{table:edu} presents the estimation results for the coefficients in the binary choice model. The method in this paper shows a positive sign of parents' education and a negative sign of being black for educational attainment. However, the HAS method shows negative signs for all coefficients, which seems inconsistent with economic intuition. Parents' education and living closer to a college are likely to increase the chance of attending a college instead of decreasing the chance. 
The results show that misspecifications in misreporting processes or distributional assumptions may lead to opposite signs of the coefficients.

\section{Extension: Two Instruments}\label{sec:exte}

Section \ref{subsec:z} and \ref{subsec:w} provide bounds for the true conditional probability $p^{*}(x)$ when there is only one instrument available.  This section studies the joint identifying power of the two instruments $(Z, W)$. The observed variables are $(X, W, Y)=(\Xz, Z, W, Y)$ when two instruments are available. 
We adjust previous assumptions in Section \ref{subsec:z} and \ref{subsec:w} slightly to accommodate the availability of the two instruments.

\begin{ass}\label{ass_zw}

\begin{enumerate}[(1)]

\item Exclusion:  for any $\xz \in \tilde{\mathcal{X}}, z\in \mathcal{Z}, w\in \mathcal{W}$, and $y\in\{0, 1\}$,  
\begin{equation*}
\begin{aligned}
\Pr(Y^{*}=1\mid x, w)&=\Pr(Y^{*}=1\mid x)=p^{*}(x), \\
\Pr(Y=1-y\mid Y^{*}=y, x, w)&=\Pr(Y=1-y\mid Y^{*}=y, \xz, w).
\end{aligned}
\end{equation*}

\item Degree of misreporting: for any $\xz \in \tilde{\mathcal{X}}$ and $w\in \mathcal{W}$,
\begin{equation*}
\Pr(Y=0\mid Y^{*}=1, \xz, w)+\Pr(Y=0\mid Y^{*}=1, \xz, w)\leq 1.
\end{equation*}

\item Monotonicity $\&$ Relevance: for any $\xz \in \tilde{\mathcal{X}}$, $w_1>w_2 \in \mathcal{W}$, and $y\in\{0, 1\}$,
 \begin{equation*}
\begin{aligned}
\Pr(Y=1-y\mid Y^{*}=y, \xz, w_1) \leq \Pr(Y=1-y\mid Y^{*}=y, \xz, w_2),  \\
\end{aligned}
\end{equation*}
and there exists $k\in\{0, 1\}$ such that the above inequality is strict.

\item Relevance: for any $\xz \in \tilde{\mathcal{X}}$, there exists $z_1\neq z_2\in \mathcal{Z}$ such that $ p^{*}(\xz, z_1)\neq p^{*}(\xz, z_2)$.

\end{enumerate}

\end{ass}

Assumption \ref{ass_zw} summarizes all assumptions for the two instruments $(Z, W)$ in previous sections. Assumption (1) states exclusion restrictions for the two instruments, which requires that instrument $Z$ does not affect misreporting probabilities and instrument $W$ does not affect the true probability. Assumption (2) requires the sum of the two-sided misreporting probabilities to be smaller than one. 
Assumption (3) adds one relevance restriction for instrument $W$ so that $W$ at least affects the misreporting probability for one group $Y^{*}=y$ strictly. Assumption (4) is the relevance condition for instrument $Z$, but the direction of how the true probability is affected by instrument $Z$ is not restricted so that instrument $Z$ can either increase or decrease the true probability. The relevance condition of instrument $Z$ can guarantee that the supremum and infimum of the reported probability over $Z$ are bounded away from one and zero respectively. Therefore, the boundary condition in previous sections is no longer needed in this section.

Under the above assumptions, we can use joint variation in the two instruments to derive bounds for misreporting probabilities and the true probability. The joint variation can bound misreporting probabilities through a new channel, thereby providing more informative results than simply taking intersections over the bounds derived using each instrument separately. 

Let $w_{m}$ denote the maximum value of instrument $W$. Next, we establish bounds for the misreporting probabilities evaluated at $w_{m}$ by using the two instruments jointly. Under the monotonicity condition of instrument $W$ in Assumption \ref{ass_zw} (iii), the misreporting probability at $w_{m}$ is the smallest misreporting probability. The bounds for misreporting probabilities evaluated at other values of $W$ can be established similarly, which will lead to the same identification result for the true probability. Therefore, we focus on the results for the smallest misreporting probabilities.

The next lemma derives bounds on misreporting probabilities evaluated at $W=w_{m}$. 

\begin{lemma} \label{lem1}
Under Assumption \ref{ass_zw}, the misreporting probability $\Pr(Y=1-y \mid Y^{*}=y, \xz, w_{m})$ can be bounded as: $[0, U_{\alpha_{y}}(\xz, w_{m})]$ for any $\xz\in \tilde{\mathcal{X}}$ and $y\in \{0, 1\}$, where 
\begin{equation*}
\begin{aligned}
&U_{\alpha_{1}}(\xz, w_{m})=1-\sup_{z, w<w_m } \left\{ \frac{q_1(\xz, w_{m}, w)\p(\xz, z_1, w)-\p(\xz, z_1, w_m)}{q_1(\xz, w_{m}, w)-1},   \p(\xz, z, w_{m})  \right\}, \\
&U_{\alpha_{0}}(\xz, w_{m})= \inf_{z, w<w_m } \left\{ \frac{q_1(\xz,w_{m}, w)\p(\xz, z_1, w)-\p(\xz, z_1, w_{m})}{q_1(\xz, w_{m}, w)-1},   \p(\xz, z, w_{m})\right\},\\
&q_1(\xz, w_{m}, w)=\frac{\p(\xz, z_1, w_{m})-\p(\xz, z_2, w_{m}) }{\p(\xz, z_1, w)-\p(\xz, z_2, w) }. 
\end{aligned}
\end{equation*}
\end{lemma}

Lemma \ref{lem1} characterizes the lower and upper bound for the misreporting probabilities evaluated at $W=w_m$ by using two instruments jointly.  The lower bounds for the smallest misreporting probabilities are zero, since we cannot rule out the possibility of no misreporting.   

The upper bounds demonstrate the joint identifying power of the two instruments.
From the definition of $U_{\alpha_{y}}(\xz, w_m)$, it uses variation from both instruments $(Z, W)$. The term $\p(\xz, z, w_m)$ shows the identifying power of instrument $Z$, and the other term involving $q_1(\xz, w_m, w)$ uses joint information of the two instruments, providing more information compared to using only instrument $W$. The main idea is that the joint variation in the two instruments imposes additional restrictions between misreporting probabilities at different values of $W$, which can further bound the misreporting probabilities.

Given bounds on misreporting probabilities, the next proposition characterizes the identification result for the true probability $p^{*}(x)$.
 
\begin{prop}\label{prop_zw}
Under Assumption \ref{ass_zw}, the true conditional choice probability $p^{*}(x)$ is bounded as $p^{*}(x)=[L_3(x), U_3(x)]$ for any $x\in \mathcal{X}$, where
\begin{equation*}
L_3(x)= \frac{\p(x, w_m)-U_{\alpha_0}(\xz, w_m)}{1-U_{\alpha_0}(\xz, w_m)},      \qquad          U_3(x)=  \frac{\p(x, w_m)}{ 1-U_{\alpha_1}(\xz, w_m)}.
\end{equation*}
And the above bounds are sharp when instrument $W$ is binary.
\end{prop}

Proposition \ref{prop_zw} establishes bounds for the true probability by using two instruments jointly, and these bounds have exhausted all possible information from assumptions and observed data when instrument $W$ is binary. 
From the definition of the bounds, the lower bound $L_3(x)$ decreases with respect to the bound $U_{\alpha_0}(\xz, w_m)$ on misreporting probabilities, and the upper bound $U_3(x)$ increases with respect to $U_{\alpha_1}(\xz, w_m)$. Therefore, a smaller bound $U_{\alpha_y}(\xz, w_m)$  for misreporting probabilities would imply tighter bounds for the true conditional probability.
 As discussed, the upper bound $U_{\alpha_y}(\xz, w_m)$ on misreporting probabilities shown in Lemma \ref{lem1} would be smaller than the one by only using one instrument. Therefore, Proposition \ref{prop_zw} derives more informative bounds for $p^{*}(x)$ by using the two instruments jointly.

\section{Conclusion}\label{sec:conc}
This paper provides partial identification of various binary choice models with misreported dependent variables, including parametric, semiparametric, and panel binary choice models.  We introduce two distinct approaches by exploiting the availability of different instrumental variables, respectively.  Moreover, our approach can accommodate additional restrictions on misreporting, such as one-sided misreporting, bounded misreporting probabilities, and monotone misreporting probabilities. 



Our approach allows for flexible misreporting processes in the sense that we do not impose any parametric model for misreporting processes and allow for heterogeneous misreporting probabilities. It would be interesting to explore how additional parametric structures on misreporting can tighten the bounds for the conditional expectation of the true dependent variable. 
Furthermore, we focus on binary choice models in this paper, while the approach for handling misreporting may be applied more broadly. It still requires substantial future work to investigate how the method can be applied in other models with potential misreporting, such as ordered and multinomial choice models with misreported dependent variables.

\bibliography{misreporting}


\appendix

\section{Appendix}
We first introduce the notation for conditional misreporting probabilities. Define the conditional misreporting probability $\alpha_y(x)$ for people with $Y^{*}=y$ as follows: for $y\in\{0,1\}$,
\begin{equation*}
\alpha_y(x)=\Pr(Y=1-y\mid Y^{*}=y, X= x).
\end{equation*}

With a slight abuse of notation, we use the same function name $\alpha_y$ when it is conditional on different covariates to avoid the complexity of introducing more notation. As such, the function $\alpha_y$ is defined conditional on $\xz$ in Section \ref{subsec:z}, conditional on $(x, w)$ in Section \ref{subsec:w}, and conditional on $(\xz, w)$ in Section \ref{sec:exte}.

\subsection{Proof of Proposition \ref{prop_z} } \label{proof_z}
\begin{proof}

The covariate $X$ is divided into two parts: $X=(\Xz, Z)$. Under the exclusion restriction of instrument $Z$ (Assumption \ref{exc1}), we know that the conditional misreporting probability only depends on covariate $\xz$ so it is denoted as $\alpha_y(\xz)$ for $y\in\{0, 1\}$.

The proof of Proposition \ref{prop_z} comprises two steps: the first step is to bound the misreporting probabilities $\alpha_0(\xz), \alpha_1(\xz)$ using variation in instrument $Z$. The second step is to bound the true probability $p^{*}(x)$. 

\textbf{Step 1}:  bound the misreporting probabilities $\alpha_y(\xz)$. The reported probability $p(x)=\Pr(Y=1\mid x)$  comes from two parts: people with $Y^{*}=1$ and report the truth, as well as people with $Y^{*}=0$ and misreport. Then under Assumption \ref{exc1}, the reported probability $p(x)$ can be expressed as follows:
\begin{equation*}
p(x)=[1-\alpha_1(\xz)]p^{*}(x)+\alpha_0(\xz)[1-p^{*}(x)].
\end{equation*}

By combining the common term $p^{*}(x)$, the above equation can be written as
\begin{equation*}
[1-\alpha_0(\xz)-\alpha_1(\xz)]p^{*}(x)=p(x)-\alpha_0(\xz).
\end{equation*}

Assumption \ref{deg1} (degree of misreporting) implies that $ 1-\alpha_0(\xz)-\alpha_1(\xz)\in[ 0, 1]$. 
The fact that the true conditional probability satisfies $p^{*}(x)\in[0, 1]$ can bound the misreporting probability $\alpha_y(\xz)$ as follows:
\begin{equation*}
\begin{aligned}
p^{*}(x)\geq 0 \ \forall z &\Longrightarrow 0 \leq \alpha_0(\xz)\leq  p(x), \\
p^{*}(x)\leq 1 \ \forall z &\Longrightarrow  0 \leq \alpha_1(\xz) \leq 1-p(x).
\end{aligned}
\end{equation*}

Since the misreporting probability $\alpha_y(\xz)$ does not depend on $z$, we can take the smallest upper bound over $z$:
\begin{equation*}
\begin{aligned}
&0\leq \alpha_0(\xz)\leq  \inf_{z\in \mathcal{Z}}p(\xz, z)=  \under{p}_z(\xz), \\
&0 \leq \alpha_1(\xz) \leq 1- \sup_{z\in \mathcal{Z}} p(\xz, z)=1-\bar{p}_z(\xz).
  \end{aligned}
\end{equation*} 

\textbf{Step 2}: bound the true probability  $p^{*}(x)$.   Now we revisit the equation for the  reported probability $p(x)$:
\begin{equation}\label{rep}
p(x)=[1-\alpha_1(\xz)]p^{*}(x)+\alpha_0(\xz)[1-p^{*}(x)].
\end{equation}

Given bounds on misreporting probabilities $\alpha_0(\xz), \alpha_1(\xz)$ derived in the first step, equation \eqref{rep} leads to the following inequalities:
\begin{equation*}
\begin{aligned}
p(x)&=[1-\alpha_1(\xz)]p^{*}(x)+\alpha_0(\xz) [1-p^{*}(x)]  \geq \bar{p}_z(\xz)p^{*}(x), \\
p(x)&=[1-\alpha_1(\xz)]p^{*}(x)+\alpha_0(\xz) [1-p^{*}(x)]   \leq p^{*}(x)+\under{p}_z(\xz) [1-p^{*}(x)].
\end{aligned}
\end{equation*}

Under Assumption \ref{bou1} (boundary condition), we know that the infimum and supremum of the reported probability $p(x)$ is bounded away from one and zero respectively.
 Then the true probability $p^{*}(x)$ can be bounded as follows:
\begin{equation*}
\frac{p(x)-\under{p}_z(\xz) }{1-  \under{p}_z(\xz) }   \leq p^{*}(x) \leq \frac{p(x)}{\bar{p}_z(\xz) }.
\end{equation*}

Now we need to prove the sharpness of the above bounds. It can be proved by showing that the lower bound and upper bound can be achieved. The idea is to show that given the lower bound and upper bound, we can construct misreporting probabilities and the true probability which match the reported probability $p(x)$ and satisfy Assumptions \ref{exc1}-\ref{bou1}.

We first look at the upper bound. The misreporting probability is constructed as $\alpha_1(\xz)=1-\bar{p}_z(\xz)$, $\alpha_0(\xz)=0$, and the true probability is constructed as $p^{*}(x)=\frac{p(x)}{\bar{p}_z(\xz) }$. It can be verified that this construction matches the reported probability $p(x)$ and satisfies assumptions. Similarly the lower bound can be achieved when $\alpha_1(\xz)=0$, $\alpha_0(\xz)=\under{p}_z(\xz) $, and $p^{*}(x)=\frac{p(x)-\under{p}_z(\xz)  }{1- \under{p}_z(\xz) }$.


%

\end{proof}

\subsection{Proof of Proposition \ref{prop_w} } \label{proof_w}

\begin{proof}
In this part, the misreporting probability depends on the covariate $(x, w)$ so it is denoted as $\alpha_y(x, w)$. 
The strategy is similar to the proof in Section \ref{proof_z}: we first establish bounds for the misreporting probability $\alpha_y(x, w)$ using instrument $W$ and then derive bounds for the true probability $p^{*}(x)$. 

Under Assumption \ref{exc2} (exclusion) for instrument $W$, the reported probability $\p(x, w)$ can be expressed as
\begin{equation*}
\p(x, w)=[1-\alpha_1(x, w)]p^{*}(x)+\alpha_0(x, w)[1-p^{*}(x)].
\end{equation*}

Given the fact that the true probability satisfies $p^{*}(x)\in[0, 1]$ and the sum of two-sided misreporting probabilities is smaller than one (Assumption \ref{bou2}), the misreporting probabilities are bounded as
\begin{equation}\label{m2}
\begin{aligned}
p^{*}(x)\geq 0 \  &\Longrightarrow 0\leq \alpha_0(x, w)\leq  \p(x, w), \\
p^{*}(x)\leq 1 \  &\Longrightarrow    0\leq \alpha_1(x, w) \leq 1-\p(x, w).
\end{aligned}
\end{equation}

However the misreporting probability $\alpha_y(x, w)$ also depends on instrument $W$ so that the above results are not informative for the true probability. Next we use the monotonicity assumption of instrument $W$ to further bound the misreporting probabilities. Under Assumption \ref{mono} (monotonicity), the following holds for any $\tilde{w}<w\in \mathcal{W}$:
\begin{equation}
\begin{aligned}
&\alpha_0(x ,w)\leq \alpha_0(x, \tilde{w}) \leq \p(x, \tilde{w}),  \\
&\alpha_1(x ,w)\leq \alpha_1(x, \tilde{w}) \leq 1-\p(x, \tilde{w}).  \\
\end{aligned}
\end{equation}

Then the misreporting probability $\alpha_y(w)$ at each $w$ can be further bounded by taking infimum over all upper bounds of misreporting probabilities evaluated at $\tilde{w}<w$:
\begin{equation*}
\begin{aligned}
&0\leq \alpha_0(x, w)\leq \inf_{\tilde{w}\leq w} \p(x, \tilde{w})=\under{p}_w(x, w), \\
&0 \leq \alpha_1(x, w) \leq 1-\sup_{\tilde{w}\leq w} \p(x,  \tilde{w})=1-\bar{p}_w(x, w).
\end{aligned}
\end{equation*}

Now we are ready to derive bounds on the true probability by the reported probability $\p(x, w)$. Given bounds on the misreporting probability $\alpha_y(x, w)$, it has the following implication for each $w$:
\begin{equation*}
\begin{aligned}
\p(x, w)&=[1-\alpha_1(x, w)]p^{*}(x)+\alpha_0(x, w) [1-p^{*}(x)]  \geq \bar{p}_w(x, w) p^{*}(x), \\
\p(x, w)&=[1-\alpha_1(x, w)]p^{*}(x)+\alpha_0(x, w) [1-p^{*}(x)]   \leq p^{*}(x)+\under{p}_w(x, w) [1-p^{*}(x)].
\end{aligned}
\end{equation*}

By Assumption \ref{bou2} (boundary condition), the reported probability is bounded away from zero and one. Then the true probability $p^{*}(x)$ can be bounded as follows: 
\begin{equation*}
\frac{\p(x, w)- \under{p}_w(x, w) }{1- \under{p}_w(x, w) }  \leq p^{*}(x)\leq \frac{\p(x, w)}{ \bar{p}_w(x, w) }.
\end{equation*}

Since the above bounds hold for any $w$ and the true probability does not depend on $w$, we can take intersections over all possible values of $w$: 
\begin{equation*}
 \sup_{w\in \mathcal{W}} \left\{  \frac{\p(x, w)- \under{p}_w(x, w) }{1- \under{p}_w(x, w) } \right\}  \leq p^{*}(x)\leq  \inf_{w\in \mathcal{W}} \left\{ \frac{\p(x, w)}{\bar{p}_w(x, w)} \right\}.
\end{equation*}

In the end, we need to show that the above bounds are sharp when instrument $W\in\{w_1,w_2\}$ only takes two values with $w_1>w_2$.  
When instrument $W$ only takes two values, the upper bound becomes $p^{*}(x) =\min \left \{ \frac{\p(x, w_1)}{\bar{p}_w(x, w_1) }, 1 \right\} =\frac{\p(x, w_1)}{\bar{p}_w(x, w_1) } $.
This bound can be achieved when  the misreporting probability satisfies  $\alpha_1(x, w_1)=1-\bar{p}_w(x, w_1), \alpha_0(x, w_1)=0$ as well as $\alpha_1(x, w_2)=1-\p(x, w_2), \alpha_0(x, w_2)=\p(x, w_2)$. It can be verified that this construction satisfies Assumptions \ref{mono}-\ref{bou2} and matches the reported probability $\p(x, w)$.

The lower bound for the true probability is $p^{*}(x)=\max\left\{ \frac{\p(x, w_1)-\under{p}_w(x, w_1)}{1-\under{p}_w(x, w_1)}, 0 \right\}=\frac{\p(x, w_1)-\under{p}_w(x, w_1)}{1-\under{p}_w(x, w_1)}$. It can be achieved when $\alpha_1(x, w_1)=0, \alpha_0(x, w_1)=\under{p}_w(x, w_1)$ and $\alpha_1(x, w_2)=1-\alpha_0(x, w_2), \alpha_0(x, w_2)=\p(x, w_2)$. This construction satisfies Assumptions \ref{mono}-\ref{bou2} and matches the reported probability.

\end{proof}

\subsection{Proof of Proposition \ref{prop_one}}

\begin{proof}
We focus on one type of one-sided misreporting where $\a_0(\xz)=0$ or $\a_0(\xz, w)=0$, and the same analysis applies to the other case.

When instrument $Z$ is available, the upper bound $U_1(x)$ remains the same and it is sharp. It can be achieved when $\alpha_1(\xz)=1-\bar{p}_z(\xz)$, $\alpha_0(\xz)=0$, and the true probability is $p^{*}(x)=\frac{p(x)}{\bar{p}_z(\xz) }=U_1(x)$. The lower bound can be established as follows:
\begin{equation*}
p(x)=[1-\alpha_1(\xz)]p^{*}(x)\leq p^{*}(x).
\end{equation*}

Now we show that the lower bound is sharp. We can construct misreporting probabilities as $\alpha_1(\xz)=0, \a_0(\xz)=0$ and the true probability as $p^{*}(x)=p(x)$ such that they match with the reported probability $p(x)$. 

When instrument $W$ is available,  one-sided misreporting assumption implies the monotonicity of $\p(x, w)$ in $w$ under the monotonicity assumption of $W$ in Assumption \ref{mono}, and the upper bound should equal one $U_2(x)=1$. This can be achieved when $1-\alpha_1(x, w)=\p(x, w)$ and $p^*(x)=1$.

For the lower bound, the following restriction holds under the exclusion restriction:
\begin{equation*}
\p(x, w)=[1-\alpha_1(x, w)]p^{*}(x)\leq p^{*}(x).
\end{equation*}

Then the true probability can be bounded by all values of $W$:
\begin{equation*}
p^{*}(x)\geq \sup_w \p(x, w).
\end{equation*}

The lower bound can be achieved when  $1-\alpha_1(x, w)=\frac{p(x, w)} {\sup_w \p(x, w) }$  and $p^{*}(x)=\sup_w \p(x, w)$, which satisfies all assumptions and matches with the reported probability. 

\end{proof}

\subsection{Proof of Propositions \ref{prop:bounds_mis} and \ref{prop:mono_mis}}
\begin{proof}

The proofs of Propositions \ref{prop:bounds_mis} and \ref{prop:mono_mis} are similar to Appendix \ref{proof_z}-\ref{proof_w}. We focus on the scenario where instrument $Z$ is available, and the analysis for instrument $W$ is omitted since the idea is the same.

We first study Proposition \ref{prop:bounds_mis} where misreporting probabilities are bounded. As shown in \ref{proof_z}, under Assumptions \ref{exc1}-\ref{bou1}, the misreporting probability $\alpha_y(\xz)$ can be bounded below:
\begin{equation*}
\begin{aligned}
\alpha_0(\xz)\in [0,  \under{p}_z(\xz)] \quad  \alpha_1(\xz)\in [0, 1-\bar{p}_z(\xz) ]. 
  \end{aligned}
\end{equation*} 

Combining Assumption \ref{ass:bounds_mis}, we can tighten the bounds for $\alpha_y(\xz)$:
\[
\alpha_0(\xz) \in [0, \min\{ \under{p}_z(\xz), \bar{\a}_0\}]\quad \alpha_1(\xz)\in [0, 1-\max\{ \bar{p}_z(\xz), 1-\bar{\a}_1\}].
\]

Recall that the reported probability $p(x)$ is given as
\[
p(x)=[1-\alpha_1(\xz)]p^{*}(x)+\alpha_0(\xz)[1-p^{*}(x)],
\]
yielding the following inequalities:
\begin{equation*}
\begin{aligned}
p(x)&=[1-\alpha_1(\xz)]p^{*}(x)+\alpha_0(\xz) [1-p^{*}(x)]  \geq \max\{ \bar{p}_z(\xz), 1-\bar{\a}_1\} p^{*}(x), \\
p(x)&=[1-\alpha_1(\xz)]p^{*}(x)+\alpha_0(\xz) [1-p^{*}(x)]   \leq p^{*}(x)+ \min\{ \under{p}_z(\xz), \bar{\a}_0\} [1-p^{*}(x)].
\end{aligned}
\end{equation*}

Therefore, the true choice probability $p^{*}(x)$ can be bounded as
\begin{equation*}
\frac{p(x)- \min \left\{ \under{p}(\xz), \bar{\a}_0 \right\}}{1-  \min \left\{ \under{p}(\xz), \bar{\a}_0 \right\} }   \leq p^{*}(x) \leq \frac{p(x)}{\max \left\{ \bar{p}(\xz), 1-\bar{\a}_1 \right\} }.
\end{equation*}

In Proposition \ref{prop:mono_mis} with monotone misreporting probabilities $\a_0(\xz)\leq \a_1(\xz)$, the misreporting probabilities can be further bounded as
\begin{equation*}
\begin{aligned}
\alpha_0(\xz)\in [0,  \min\{ \under{p}_z(\xz), 1-\bar{p}_z(\xz)\} ] \quad  \alpha_1(\xz)\in [0, 1-\bar{p}_z(\xz) ]. 
  \end{aligned}
\end{equation*} 

Under similar arguments, we have
\begin{equation*}
\begin{aligned}
p(x)&=[1-\alpha_1(\xz)]p^{*}(x)+\alpha_0(\xz) [1-p^{*}(x)]  \geq \bar{p}_z(\xz)  p^{*}(x), \\
p(x)&=[1-\alpha_1(\xz)]p^{*}(x)+\alpha_0(\xz) [1-p^{*}(x)]   \leq p^{*}(x)+  \min\{ \under{p}_z(\xz), 1-\bar{p}_z(\xz)\}  [1-p^{*}(x)].
\end{aligned}
\end{equation*}

Therefore, misreporting probabilities can be bounded as
\[
\frac{p(x)-  \min\left\{ \under{p}_z(\xz), 1-\bar{p}_z(\xz) \right\} }{1-  \min\left\{ \under{p}_z(\xz), 1-\bar{p}_z(\xz) \right\} }   \leq p^{*}(x) \leq \frac{p(x)}{ \bar{p}(\xz) }.
\]

\end{proof}

\subsection{Proof of Lemma \ref{lem1} }

\begin{proof}
In this part, the misreporting probability $\alpha_y(\xz, w)$ depends on $(\xz, w)$. We suppress the covariate $\Xz$ in this proof to simplify notation.
Under the exclusion restrictions imposed on the two instruments in Assumption \ref{ass_zw}, the reported probability $\p(z, w)$ can be expressed as follows:
 \begin{equation}\label{two_exc}
\p(z, w)=[1-\alpha_1(w)] p^{*}(z)+\alpha_0(w) [1- p^{*}(z)].
\end{equation}

Under Assumption \ref{ass_zw}, instrument $Z$ only affects the true probability $p^{*}(z)$ and instrument $W$ only affects the misreporting probabilities $\alpha_y(w)$. We first derive bounds on misreporting probabilities $\alpha_y(w)$ and then establish bounds for the true probability based on equation \eqref{two_exc}. 

We first look at identification for the misreporting probabilities $\alpha_y(w)$.  Since the true conditional probability satisfies $p^{*}(z)\in[0, 1]$ for any $z$,  we can bound $\alpha_y(w)$ following similar arguments to the proofs in Section \ref{proof_z} and \ref{proof_w}:
\begin{equation}\label{b1}
\alpha_0(w) \in [0,  \inf_{z\in \mathcal{Z}} \p(z, w)], \qquad \alpha_1(w)\in [0, 1-\sup_{z\in \mathcal{Z}}  \p(z, w)].
\end{equation}

Next, we use joint variation in the two instruments to build relationships between misreporting probabilities evaluated at different values of $W$. This relationship can further bound the misreporting probabilities. We fix instrument $Z=z$ and look at the reported probability evaluated at two different values  $w_1\neq w_2\in \mathcal{W}$:
\begin{align*}
\p(z, w_1)&=[1-\alpha_1(w_1)] p^{*}(z)+\alpha_0(w_1) [1- p^{*}(z)], \\
\p(z, w_2)&=[1-\alpha_1(w_2)] p^{*}(z)+\alpha_0(w_2) [1- p^{*}(z)].
\end{align*}

From condition \eqref{b1}, we know that $1-\alpha_1(w)-\alpha_0(w)\geq \sup\limits_{z\in \mathcal{Z}}  \p(z, w)-\inf\limits_{z\in \mathcal{Z}}  \p(z, w)>0$ by the relevance condition of instrument $Z$. The above two equations both contain the common term $p^{*}(z)$, and canceling out the same term $p^{*}(z)$ has the following implication:
\begin{equation*}
\p(z, w_1)=\frac{1-\alpha_{1}(w_1)-\alpha_{0}(w_1)}{1-\alpha_{1}(w_2) -\alpha_{0}(w_2)} [\p(z, w_2)-\alpha_0(w_2)]+\alpha_0(w_1).
\end{equation*}

Let $A_1(w_1, w_2)=\frac{1-\alpha_{1}(w_1)-\alpha_{0}(w_1)}{1-\alpha_{1}(w_2) -\alpha_{0}(w_2)}$ and $A_0(w_1, w_2)=\alpha_0(w_2)A_1(w_1, w_2)-\alpha_0(w_1)$. Then the above equation can be rewritten as follows for $z\in \{z_1, z_2\}$:
\begin{align*}
\p(z_1, w_1)&=A_1(w_1, w_2) \p(z_1, w_2)-A_0(w_1, w_2), \\
\p(z_2, w_1)&=A_1(w_1, w_2) \p(z_2, w_2)-A_0(w_1, w_2).
\end{align*}

The two equations can jointly identify  $A_1(w_1, w_2)$ and $A_0(w_1, w_2)$ as long as the equations are not collinear. By the relevance condition of instrument $Z$, we know that $\p(z_1, w)-\p(z_2, w)=[1-\alpha_1(w)-\alpha_0(w)][p^{*}(z_1)-p^{*}(z_2)]\neq 0$ so the two equations are not collinear. Then $A_1(w_1, w_2)$ and $A_0(w_1, w_2)$ can be  identified as follows:
\begin{align*}
A_1(w_1, w_2)&=\frac{\p(z_1, w_1)-\p(z_2, w_1) }{\p(z_1, w_2)-\p(z_2, w_2) }\equiv q_1(w_1, w_2), \\
A_0(w_1, w_2)&=q_1(w_1, w_2)\p(z, w_2)-\p(z, w_1)\equiv q_0(w_1, w_2) \quad  \forall z.
\end{align*}

According to the definition of  $A_1(w_1, w_2)$ and $A_0(w_1, w_2)$, they build relationships between misreporting probabilities and this relationship can be used to further bound misreporting probabilities. From the definition of $A_1(w_1, w_2)$ and $A_0(w_1, w_2)$, the following holds for $\alpha_y(w)$: 
\begin{equation}\label{con}
\begin{aligned}
1-\alpha_1(w_1)&= [1-\alpha_1(w_2)]q_1(w_1, w_2)-q_0(w_1, w_2),      \\
\alpha_0(w_1)&=\alpha_0(w_2)q_1(w_1, w_2)-q_0(w_1, w_2).
\end{aligned}
\end{equation}

By the monotonicity condition of instrument $W$ in Assumption \ref{ass_zw} and bounds on misreporting probabilities in condition \eqref{b1}, the following condition summarizes restrictions on misreporting probabilities:
\begin{equation}\label{res}
\begin{aligned}
1 \geq 1-\alpha_1(w_1)\geq 1-\alpha_1(w_2),& \qquad  0 \leq \alpha_0(w_1)  \leq  \alpha_0(w_2), \\
1-\alpha_1(w) \geq \sup_{z\in \mathcal{Z}} \p(z, w),&  \qquad  \alpha_0(w) \leq \inf_{z\in \mathcal{Z}} \p(z, w).
\end{aligned}
\end{equation}

Given the above restrictions on $\alpha_y(w)$ and equation \eqref{con}, we can derive bounds on misreporting probabilities $\alpha_y(w_1)$. In order to derive explicit bounds on $\alpha_y(w_1)$,  we also need to discuss the value $q_1(w_1, w_2)$ and $q_0(w_1, w_2)$. By the monotonicity condition of instrument $W$, the following holds:
\begin{equation*}
\begin{aligned}
q_1(w_1, w_2)&=\frac{1-\alpha_{1}(w_1)-\alpha_{0}(w_1)}{1-\alpha_{1}(w_2) -\alpha_{0}(w_2)}> 1, \\
q_0(w_1, w_2)&= \alpha_0(w_2)q_1(w_1, w_2)-\alpha_0(w_1)\geq \alpha_0(w_1)[q_1(w_1, w_2)-1] \geq 0, \\
q_1(w_1, w_2)-q_0(w_1, w_2)-1 &=q_1(w_1, w_2)-\alpha_1(w_1)-[1-\alpha_1(w_2)]q_1(w_1, w_2) \\
&=\alpha_1(w_2)q_1(w_1, w_2) -\alpha_1(w_1)\geq 0.
\end{aligned}
\end{equation*}

Then conditions \eqref{con} together with the restrictions \eqref{res} leads to bounds for $\alpha_y(w_1)$ for any $w_1, w_2$:
\begin{equation*}
\begin{aligned}
\max \left\{ \frac{q_0(w_1, w_2)}{q_1(w_1, w_2)-1},  \sup_{z\in \mathcal{Z}} \p(z, w_1)  \right\}&\leq 1-\alpha_1(w_1) \leq 1, \\
0 &\leq \alpha_0(w_1) \leq \min \left\{ \frac{q_0(w_1, w_2)}{q_1(w_1, w_2)-1},  \inf_{z\in \mathcal{Z}} \p(z, w_1)  \right\}.
\end{aligned}
\end{equation*}

The above bounds hold for any $w_2<w_1$ and any $z$. Therefore we can derive bounds for $\alpha_y(w_m)$ by taking intersections over bounds derived from all possible values of $z, w<w_m$ which leads to:
\begin{equation*}
\begin{aligned}
0\leq \alpha_1(w_m)&\leq 1- \sup_{z\in \mathcal{Z}, w<w_m\in \mathcal{W}} \left\{ \frac{q_0(w_m, w)}{q_1(w_m, w)-1},   \p(z, w_m)  \right\}=U_{\alpha_1}(w_m),  \\
 0\leq \alpha_0(w_m)& \leq \inf_{z\in \mathcal{Z}, w<w_m\in \mathcal{W}} \left\{ \frac{q_0(w_m, w)}{q_1(w_m, w)-1},   \p(z, w_m)  \right\}=U_{\alpha_0}(w_m).
\end{aligned}
\end{equation*}

 \end{proof}

 \subsection{Proof of Proposition \ref{prop_zw} }
 \begin{proof}

We look at  the equation for the reported probability conditional on $(x, w_m)$:
 \begin{equation}
\p(x, w_m)=[1-\alpha_1(\xz, w_m)] p^{*}(x)+\alpha_0(\xz, w_m) [1- p^{*}(x)].
\end{equation}

Given bounds on misreporting probabilities derived in Lemma \ref{lem1}, it implies the following conditions:
\begin{equation*}
\begin{aligned}
\p(x, w_m)&\leq p^{*}(x)+U_{\alpha_0}(\xz, w_m)[1-p^{*}(x)], \\
\p(x, w_m)&\geq [1-U_{\alpha_1}(\xz, w_m)] p^{*}(x).
\end{aligned}
\end{equation*}

First we can show that the upper bounds for $U_{\alpha_1}(\xz, w_m)$ and $U_{\alpha_0}(\xz, w_m)$  are strictly smaller than one. It can be shown by proving that  $\sup_z p(\xz, z, w)>0$ and $\inf_z \p(\xz, z, w)<1$ for any $(\xz, w)$. 

We prove it by contradiction. If the supremum of $\p(\xz, z, w)$ over $z$ is zero which implies that $\p(\xz, z, w)=0$ for all $z$.  However we know that there exists $z_1, z_2$ such that $\p(\xz, z_1, w)-\p(\xz, z_2, w)=[1-\alpha_1(\xz, w)-\alpha_0(\xz, w)] [p^{*}(\xz, z_1)-p^{*}(\xz, z_2)]\neq 0$ by the relevance condition of instrument $Z$ and the degree of misreporting assumptions in Assumption \ref{ass_zw}. Therefore $\p(\xz, z, w)$ cannot be zero for all $z$ and the supremum of it over $z$ is strictly larger than zero. Similarly we can conclude that $\inf_z \p(\xz, z, w)<1$ and the upper bounds $U_{\alpha_y}(\xz, w_m)$ are strictly smaller than one.

Then the true probability $p^{*}(x)$ can be bounded as follows:
\begin{equation*}
\frac{\p(x, w_m)-U_{\alpha_0}(\xz, w_m) }{1-U_{\alpha_0}(\xz, w_m)}  \leq p^{*}(x)\leq  \frac{\p(x, w_m)}{ 1-U_{\alpha_1}(\xz, w_m)}.
\end{equation*}

Now we need to show that the above bounds are sharp when instrument $W$ only takes two values: $W\in\{w_1, w_2\}$ with $w_1>w_2$ so that $w_m=w_1$. 
We prove it by constructing misreporting probabilities and the true probability such that they match with the reported probability and also satisfy Assumption \ref{ass_zw}. 

We first show that the upper bound $p^{*}(x)=\frac{\p(x, w_1)}{ 1-U_{\alpha_1}(\xz, w_1)}$ can be achieved. 
The misreporting probabilities $\alpha_y(\xz, w)$ are constructed as: $\alpha_1(\xz, w_1)=U_{\alpha_1}(\xz, w_1)$, $\alpha_0(\xz, w_1)=0$,  $\alpha_1(\xz, w_2)=1-\frac{1-U_{\alpha_1}(\xz, w_1)+q_0(\xz, w_1, w_2)}{q_1(\xz, w_1, w_2)}, \alpha_0(\xz, w_2)=\frac{q_0(\xz, w_1, w_2)}{q_1(\xz, w_1, w_2)}$.  It can be verified that they match with the reported probability $\p(x, w)$ for any $x, w\in\{ w_1, w_2\}$ under some algebra.

Next we verrity that misreporting probabilities we construct satisfy Assumption \ref{ass_zw} which assumes monotonicity and the degree of misreporting. Given the upper bound $U_{\alpha_1}(\xz, w_1)$ is strictly smaller than one, then the misreporting probabilities constructed above satisfy the degree of misreporting: $\alpha_0(\xz, w)+\alpha_1(\xz, w)<1$ for any $\xz, w\in\{w_1, w_2\}$.
The monotonicity condition for $\alpha_0(\xz, w)$ is satisfied since $\alpha_0(\xz, w_2)\geq 0=\alpha_0(\xz, w_1)$, so we only need to show the monotonicity condition for $\alpha_1(\xz, w)$. We look at the difference of the two misreporting probabilities multiplied by $q_1(\xz, w_1, w)$:
\begin{equation*}
\begin{aligned}
&q_1(\xz, w_1, w_2)[\alpha_1(\xz, w_1)-\alpha_1(\xz, w_2)] \\
=&-[1-U_{\alpha_1}(\xz, w_1)][q_1(\xz, w_1, w_2)-1]+q_0(\xz, w_1, w_2) \\
\leq &-\frac{q_0(\xz, w_1, w_2)}{q_1(\xz, w_1, w_2)-1}[q_1(\xz, w_1, w_2)-1]+q_0(\xz, w_1, w_2)\leq 0.
\end{aligned}
\end{equation*}
Therefore the monotonicity condition is also satisfied. 

Lastly, we need to prove that the lower bound $p^{*}(x)=\frac{\p(x, w_1)-U_{\alpha_0}(\xz, w_1) }{1-U_{\alpha_0}(\xz, w_1)}$ can be obtained. The misreporting probabilities are constructed as: $\alpha_1(\xz, w_1)=0$, $\alpha_0(\xz, w_1)=U_{\alpha_0}(\xz, w_1)$,  $\alpha_1(\xz, w_2)=1-\frac{1+q_0(\xz, w_1, w_2)}{q_1(\xz, w_1, w_2)}$, and $ \alpha_0(\xz, w_2)= \frac{U_{\alpha_0}(\xz, w_1)+q_0(\xz, w_1, w_2)}{q_1(\xz, w_1, w_2)}$.  Similarly it can be shown that they satisfy Assumption \ref{ass_zw} and match with the reported probability $\p(x, w)$ for any $(x, w)$.

\end{proof}

\end{document}